\documentclass[prd,twocolumn,showpacs,amsmath,amssymb,floatfix,superscriptaddress,nofootinbib]{revtex4}

\usepackage{graphicx}
\usepackage{color}
\usepackage[colorlinks=true, linkcolor=blue, citecolor=blue, urlcolor=blue]{hyperref}

\newcommand{\DS}[1]{/\!\!\!#1}

\allowdisplaybreaks

\begin{document}

\title{Properties of the $\eta_q$ leading-twist distribution amplitude and its effects to the $B/D^+ \to\eta^{(\prime)}\ell^+ \nu_\ell$ decays}

\author{Dan-Dan Hu}
\email{hudd@stu.cqu.edu.cn}
\author{Xing-Gang Wu}
\email{wuxg@cqu.edu.cn}
\address{Department of Physics, Chongqing Key Laboratory for Strongly Coupled Physics, Chongqing University, Chongqing 401331, P.R. China}
\author{Hai-Bing Fu}
\email{fuhb@cqu.edu.cn}
\author{Tao Zhong}
\email{zhongtao1219@sina.com}
\author{Zai-Hui Wu}
\email{wuzaihui@hotmail.com}
\address{Department of Physics, Guizhou Minzu University, Guiyang 550025, P.R. China}
\author{Long Zeng}
\email{zlong@cqu.edu.cn}
\address{Department of Physics, Chongqing Key Laboratory for Strongly Coupled Physics, Chongqing University, Chongqing 401331, P.R. China}

\date{\today}

\begin{abstract}

The $\eta^{(\prime)}$-mesons in the quark-flavor basis are mixtures of two mesonic states $|\eta_{q}\rangle=|\bar u u+\bar d d\rangle/\sqrt 2$ and $|\eta_{s}\rangle=|\bar s s\rangle$. In previous work, we have made a detailed study on the $\eta_{s}$ leading-twist distribution amplitude by using the $D^+_s$ meson semileptonic decays. As a sequential work, in the present paper, we fix the $\eta_q$ leading-twist distribution amplitude by using the light-cone harmonic oscillator model for its wave function and by using the QCD sum rules within the QCD background field to calculate its moments. The input parameters of $\eta_q$ leading-twist distribution amplitude $\phi_{2;\eta_q}$ at the initial scale $\mu_0\sim 1$ GeV are fixed by using those moments. The QCD sum rules for the $0_{\rm th}$-order moment can also be used to fix the magnitude of $\eta_q$ decay constant, giving $f_{\eta_q}=0.141\pm0.005$ GeV. As an application of $\phi_{2;\eta_q}$, we calculate the transition form factors $B(D)^+ \to\eta^{(\prime)}$ by using the QCD light-cone sum rules up to twist-4 accuracy and by including the next-to-leading order QCD corrections to the leading-twist part, and then fix the related CKM matrix element and the decay width for the semi-leptonic decays $B(D)^+ \to\eta^{(\prime)}\ell^+ \nu_\ell$.

\end{abstract}

\date{\today}


\maketitle

\section{Introduction}

The mixing of $\eta$ and $\eta'$ mesons is essential to disentangle the standard model (SM) hadronic uncertainties with the  new physics beyond the SM. It involves the dynamics and structure of the pseudoscalar mesons that has two mixing modes $\eta-\eta'$ and $\eta-\eta'-G$, both of which have important theoretical significance. These mixings are caused by the QCD anomalies and are related to the breaking of chiral symmetry. However, since the matrix element of the exception operator is mainly non-perturbative, it still has not been calculated reliably. One may turn to phenomenological studies to obtain useful information on the non-perturbative QCD theory~\cite{Ke:2010htz, Ke:2011fj, Cao:2012nj}. At present, the $\eta-\eta'-G$ mixing mode has been studied in detail in Refs~\cite{Gilman:1987ax, Ball:1995zv, Feldmann:2002kz, Kroll:2002nt, Ambrosino:2009sc, Ball:2007hb, Duplancic:2015zna}. As for the $\eta-\eta'$ mixing model, one can investigate it by using two distinct schemes, namely the singlet-octet (SO) scheme and the quark-flavor (QF) scheme. These two schemes reflect different understandings of the essential physics and they are related with a proper rotation of an ideal mixing angle~\cite{Cao:2012nj}. Practically, a dramatic simplification can be achieved by adopting the QF scheme~\cite{Schechter:1992iz, Feldmann:1999uf, Feldmann:1998su, Feldmann:1998vh}, especially, the decay constants in the quark-flavor basis simply follow the same pattern of the state mixing due to the OZI-rule. In the present paper, we shall adopt the QF scheme to do our analysis and to achieve a better understanding of the mixing mechanism between $\eta$ and $\eta'$.

The heavy-to-light $B(D)\to\eta^{(\prime)}$ transitions are important, since they involve $b\to u$ and $c\to d$ transitions and are sensitive to the Cabibbo-Kobayashi-Maskawa (CKM) matrix elements $|V_{\rm ub}|$ and $|V_{\rm cd}|$. A more precise determination of $|V_{\rm ub}|$ and $|V_{\rm cd}|$ would improve the stringency of unitarity constraints on the CKM matrix elements and provides an improved test of SM. Many measurements on $|V_{\rm ub}|$ and $|V_{\rm cd}|$ have been done according to various decay channels of $B(D)$-mesons~\cite{CLEO:2007vpk, BESIII:2013iro, BaBar:2014xzf, CLEO:2009svp, HFLAV:2019otj, Lubicz:2017syv, BaBar:2011xxm, Belle:2005uxj, Gonzalez-Solis:2018ooo, ParticleDataGroup:2022pth}. Compared with the non-leptonic $B(D)$-meson decays, the semi-leptonic decays $ D^+ \to\eta^{(\prime)}\ell^+ \nu_\ell$~\cite{CLEO:2008xqh, CLEO:2010pjh, BESIII:2018eom, Ablikim:2020hsc} and $ B^+ \to\eta^{(\prime)}\ell^+ \nu_\ell$~\cite{CLEO:2007vpk, BaBar:2008byi, BaBar:2010npl, Belle:2017pzx, Belle:2021hah} are much simpler with less non-perturbative effects and can serve as helpful platforms for exploring the differences among various mechanisms.

As key components of $B(D)\to\eta^{(\prime)}$ semileptonic decays, the $B(D)\to\eta^{(\prime)}$ transition form factors (TFFs) need to be precisely calculated, whose main contribution comes from the $|\eta_q\rangle$-component (the $|\eta_{s}\rangle$-component gives negligible contribution here, but will have sizable contribution for $B_s$ ($D_s$) decays~\cite{Gonzalez-Solis:2018ooo}). By further assuming the $SU_{\rm F}(3)$ symmetry, the TFFs $f_+^{B(D)\to\eta^{(\prime)}}$ in the QF scheme satisfy the relations~\cite{Gonzalez-Solis:2018ooo, Cheng:2010yd}
\begin{eqnarray}
f_+^{B(D)\to\eta}  &=& \cos \phi f_+^{B(D)\to\eta_q}, \label{Eq:ma0} \\
f_+^{B(D)\to\eta'} &=& \sin \phi f_+^{B(D)\to\eta_q},
\label{Eq:ma}
\end{eqnarray}
where $\phi$ is the mixing angle between the $|\eta_q\rangle$-component and the $|\eta_s\rangle$-component.

The TFFs of the heavy-to-light transitions at large and intermediate momentum transfers are among the most important applications of the light-cone sum rules (LCSR) approach. Using the LCSR approach, a two-point correlation function (correlator) will be introduced and expanded near the light cone $x^2 \to 0$, whose transition matrix elements are then parameterized as the light meson's light-cone distribution amplitudes (LCDAs) of increasing twists~\cite{Braun:1988qv, Balitsky:1989ry, Chernyak:1990ag, Ball:1991bs}. It is thus important to know the properties of the LCDAs.

In present paper, we adopt the light cone harmonic oscillator (LCHO) model for the $\eta_q$ leading-twist LCDA $\phi_{2;\eta_q}$. The LCHO model is based on the well-known Brodsky-Huang-Lepage (BHL) prescription~\cite{Brodsky:1981jv, Lepage:1982gd}~\footnote{The BHL-prescription is obtained via the way of connecting the equal-time wavefunction in the rest frame and the wavefunction in the infinite momentum frame, which indicates that the LCWF is a function of the meson's off-shell energy.} for the light-cone wavefunction (LCWF), which is composed of the spin-space LCWF and the spatial one. The LCDA can be obtained by integrating over the transverse momentum from the LCWF. The parameters of $\phi_{2;\eta_q}$ at an initial scale will be fixed by using the derived moments of the LCDA, which will then be run to any scale region via proper evolution equation. Its moments will be calculated by using the QCD sum rules within the framework of the background field theory (BFTSR)~\cite{Huang:1986wm, Huang:1989gv}. The QCD sum rules approach suggests to use the non-vanishing vacuum condensates to represent the non-perturbative effects~\cite{Shifman:1978bx}. The QCD background field approach provides a simple physical picture for those vacuum condensates from the viewpoint of field theory~\cite{Hubschmid:1982pa, Govaerts:1984bk, Reinders:1984sr, Elias:1987ac}. It assumes that the quark and gluon fields are composed of background fields and the quantum fluctuations around them. And the vacuum expectation values of the background fields describe the non-perturbative effects, while the quantum fluctuations represent the calculable perturbative effects. As a combination, the BFTSR approach provides a clean physical picture for separating the perturbative and non-perturbative properties of the QCD theory and provides a systematic way to derive the QCD sum rules for hadron phenomenology, which greatly simplifies the calculation due to its capability of adopting different gauges for quantum fluctuations and background fields. Till now, the BFTSR approach has been applied for dealing with the LCDAs of various mesons, some recent examples can be found in Refs.\cite{Zhang:2021wnv, Zhong:2018exo, Fu:2018vap, Zhang:2017rwz, Fu:2016yzx}.

The remaining parts of the paper are organized as follows. In Sec.~\ref{sec:2}, we give the calculation technology for deriving the moments of $\eta_q$ leading-twist LCDA $\phi_{2;\eta_q}$ by using the BFTSR approach, give a brief introduction of the LCHO model of $\phi_{2;\eta_q}$, and then give the LCSRs for the TFFs of the semi-leptonic decay $ B(D)^+ \to\eta_q\ell^+ \nu_\ell$. In Sec.~\ref{sec:3}, we first determine the parameters of $\phi_{2;\eta_q}$, and then the TFFs, the decay width and the CKM matrix element of the semi-leptonic decay $ B(D)^+\to\eta^{(\prime)}\ell^+ \nu_\ell$ will be discussed. We will also compare our results with the experimental data and other theoretical predictions. Sec.~\ref{sec:summary} is reserved for a summary.

\section{Calculation Technology}\label{sec:2}

\subsection{Determination of the moments $\langle \xi _{2;\eta_q}^n\rangle$ of the $\eta_q$ leading-twist LCDA using the BFTSR}

For the QF scheme, the physical meson states $|\eta\rangle$ and $|\eta'\rangle$ are related to the QF basis $|\eta_q\rangle=|\bar uu+\bar d d\rangle/\sqrt 2$ and $|{\eta _s}\rangle=|\bar s s\rangle$ by an orthogonal transformation~\cite{Feldmann:1998vh},
\begin{eqnarray}
\begin{pmatrix} |\eta\rangle \cr |\eta'\rangle \end{pmatrix}
&=& \begin{pmatrix} \cos\phi & -\sin\phi \cr
\sin\phi & \cos\phi \end{pmatrix} \begin{pmatrix} |\eta_q\rangle \cr |\eta_s\rangle \end{pmatrix},
\label{mixangle}
\end{eqnarray}
where $\phi$ is the mixing angle. For the QF basis, one has two independent types of axial vector currents $J_{\mu5}^q~(q=u,d)$ and $J_{\mu5}^s$, e.g.
\begin{align}
J^q_{\mu 5}=\frac{1}{\sqrt 2}(\bar u \gamma_\mu\gamma_5 u+\bar d \gamma_\mu\gamma_5 d),~~J^s_{\mu 5}=\bar s \gamma_\mu\gamma_5 s.
\end{align}
Their corresponding decay constants are
\begin{align}
\langle 0|J_{\mu 5}^{j}|\eta^{(\prime)} (p)\rangle=i f_{\eta^{(\prime)}_{j}} p_{\mu},~~~~~j=(q,s)
\label{Eq:QF}
\end{align}
where $p$ is the momentum of $\eta^{(\prime)}$. By extending the matrix elements~\eqref{Eq:QF} as the non-local operators over the light cone, one can achieve the definition of the corresponding LCDA. The LCDA of the valence quark momentum fraction distribution of $\eta^{(\prime)}$ meson can be defined similarly as those of other mesons by spreading out the non-local operators on the light cone via a way with increasing twists. In Ref.\cite{Hu:2021zmy}, the $J_{\mu5}^s$ has been adopted to study the properties of $\eta_s$, and at present, we will focus on $J_{\mu5}^q$ to study the properties of $\eta_q$.

To determine the properties of LCDA, one can firstly calculate its moments. The $\eta^{(\prime)}$ meson leading-twist LCDA are defined as~\cite{Ball:2007hb, Duplancic:2015zna}
\begin{align}
&\langle 0|\bar \Psi (z){{\rm{\cal C}}_j}[z, - z]\gamma_\mu {\gamma _5}\Psi ( - z)|\eta^{(\prime)} (p)\rangle
\nonumber\\
&= i{f_{\eta^{(\prime)}_j}}p_\mu \int\limits_0^1 {dx} {e^{i(x z \cdot p-\bar x z\cdot p)}}{\phi_{2;\eta^{(\prime)}_j}}(x), \label{Eq.6}
\end{align}
where $\Psi=(u,d,s)$ represents the triplet of light-quark fields in flavour space, $z_\mu$ stands for the light-like vector, $[z,-z]$ is the path-ordered gauge connection which ensures the gauge invariance of the operator, and ${\phi_{2;\eta^{(\prime)}_j}}(x)$ are leading-twist LCDAs of $\eta^{(\prime)}$ mesons with respect to the current whose flavour content is given by ${\rm{\cal C}}_j$ with $j=(q, s)$, respectively. We have ${\cal C}_q=(\sqrt 2{\cal C}_1+{\cal C}_8)/\sqrt 3$ and ${\cal C}_s=({\cal C}_1-\sqrt 2{\cal C}_8)/\sqrt 3$ with ${\cal C}_1={\mathbf 1}/\sqrt 3$ and ${\cal C}_8=\lambda_8/\sqrt 2$ which are derived in SO scheme~\cite{Ball:2007hb}, where $\lambda_8$ is the standard Gell-Mann matrix and ${\mathbf 1}$ is $3\times3$ unit matrix.

By doing the series expansion near $z^2 \to 0$ on both sides of Eq.(\ref{Eq.6}), one will get
\begin{align}
&\langle 0|\bar \Psi (0){{\rm{\cal C}}_j}\DS z {\gamma _5}{(iz \cdot \tensor D )^n}\Psi (0 )|\eta^{(\prime)} (p)\rangle
\nonumber\\
&= i(z \cdot p)^{n+1}{f_{\eta^{(\prime)}_j}} \int\limits_0^1 {dx} (2x-1)^n {\phi_{2;\eta^{(\prime)}_j}}(x),   \label{Eq:da}  \\
&= i(z \cdot p)^{n+1}{f_{\eta^{(\prime)}_j}} \langle \xi_{2;\eta^{(\prime)}_j}^n\rangle ,
\end{align}
where the $n_{\rm th}$-order moment of $\eta^{(\prime)}$ leading-twist LCDA $\phi_{2;\eta^{(\prime)}_j}(x)$ has been defined as
\begin{align}
\langle \xi _{2;\eta^{(\prime)}_j}^n\rangle = \int\limits_0^1 {dx} (2x-1)^n {\phi_{2;\eta^{(\prime)}_j}}(x).
\label{Eq:gm}
\end{align}

As mentioned above, the $\eta^{(\prime)}$ meson has two distinct components, $|\eta_q\rangle$ and $|\eta_s\rangle$. The $|\eta_{s}\rangle$-component has been studied by using $B_s(D_s)\to \eta^{(\prime)}$ semi-leptonic decays. Similarly, the $|\eta_{q}\rangle$-component can also be studied by using $B(D)\to \eta^{(\prime)}$ semi-leptonic decays. Eqs.(\ref{Eq:ma0},\ref{Eq:ma}) indicate that to calculate the TFFs $f_+^{B(D)\to\eta^{(\prime)}}$, we need to calculate the TFFs $f_+^{B(D)\to\eta_q}$. By further comparing theoretical predictions with the possible data on the $B(D)\to \eta^{(\prime)}$ semi-leptonic decays, we can inversely achieve useful information on the $\eta_q$ leading-twist LCDA. We will calculate the moments of the $\eta_q$ leading-twist LCDA $\phi_{2;\eta_q}(x)$, which have been defined in Eq.(\ref{Eq:gm}), within the framework of the BFTSR approach.

A recent mini-review of the basic idea and formulas for the background field theory can be found in Ref.\cite{Hu:2021zmy}. Following the standard procedures of the BFTSR approach, to derive the moments of the $\eta_q$ leading-twist LCDA, we first construct the following correlator
\begin{align}
\Pi_{2;\eta_q}^{(n,0)}(z,q) & = i\int {d^4x{e^{iq \cdot x}}} \langle 0|T\{ {J_n}(x),J_0^\dag (0)\} |0\rangle
\nonumber\\
& = {(z \cdot q)^{n + 2}}\Pi _{2;\eta_q}^{(n,0)}(q^2),
\label{Eq:CF}
\end{align}
where the currents $J_n(x) = \frac{{\cal C}_q}{\sqrt 2} [\bar u(x)\DS z\gamma_5 {(iz \cdot \tensor D )^n}u(x) \!+\! \bar d(x) \DS z\gamma_5 {(iz \cdot \tensor D )^n}d(x)]$ and $J_0^\dag (0)=\frac{{\cal C}_q}{\sqrt 2} [\bar u(0)\DS z\gamma_5 u(0)\!+\! \bar d(0) \DS z\gamma_5 d(0) ]$ with $z^2=0$. Only the even moments are non-zero due to the $G$-parity, which indicate that $n=(0,2,4,\ldots)$, respectively.

Secondly, one can apply the operator product expansion (OPE) to deal with the correlator in the deep Euclidean region. In deep Euclidean region $q^2\ll 0$, after applying the OPE, the correlator~\eqref{Eq:CF} becomes
\begin{align}
\Pi _{2;\eta_q}^{(n,0)}&(z,q)=i\int d^4xe^{iq\cdot x}\frac{1}{2}{\rm Tr}[{\rm{\cal C}}_q{\rm{\cal C}}_q]
\times \nonumber\\
&\big\{-2{\rm Tr}\langle 0|S^u_F(0,x)\DS z\gamma_5(iz \cdot \tensor D )^n S^d_F(x,0)\DS z\gamma_5|0\rangle
\nonumber\\
& ~~~+2{\rm Tr}\langle 0|\bar u(x)u(0)\DS z\gamma_5(iz \cdot \tensor D )^n S^d_F(x,0)\DS z\gamma_5|0\rangle
\nonumber\\
& ~~~+2{\rm Tr}\langle 0|S^u_F(0,x)\DS z\gamma_5(iz \cdot \tensor D )^n \bar d(0)d(x)\DS z\gamma_5|0\rangle
\nonumber\\
&~~~-{\rm Tr}\langle 0|S^u_F(0,x)\DS z\gamma_5(iz \cdot \tensor D )^n S^u_F(x,0)\DS z\gamma_5|0\rangle
\nonumber\\
& ~~~+{\rm Tr}\langle 0|\bar u(x)u(0)\DS z\gamma_5(iz \cdot \tensor D )^n S^u_F(x,0)\DS z\gamma_5|0\rangle
\nonumber\\
& ~~~+{\rm Tr}\langle 0|S^u_F(0,x)\DS z\gamma_5(iz \cdot \tensor D )^n \bar u(0)u(x)\DS z\gamma_5|0\rangle
\nonumber\\
&~~~-{\rm Tr}\langle 0|S^d_F(0,x)\DS z\gamma_5(iz \cdot \tensor D )^n S^d_F(x,0)\DS z\gamma_5|0\rangle
\nonumber\\
& ~~~+{\rm Tr}\langle 0|\bar d(x)d(0)\DS z\gamma_5(iz \cdot \tensor D )^n S^d_F(x,0)\DS z\gamma_5|0\rangle
\nonumber\\
& ~~~+{\rm Tr}\langle 0|S^d_F(0,x)\DS z\gamma_5(iz \cdot \tensor D )^n \bar d(0)d(x)\DS z\gamma_5|0\rangle
\nonumber\\
&~~~+\cdots\big\},
\end{align}
in which ${\rm Tr}[{\rm{\cal C}}_q{\rm{\cal C}}_q]=1$, $(iz \cdot \tensor D )^n$ stands for the vertex operators, $S^{u(d)}_F(0,x)$ and $S^{u(d)}_F(x,0)$ represent the $u$- and $d$- quark propagators move from $0\to x$ and $x\to 0$, respectively. The right-hand side of the correlator is perturbatively calculable within the framework of BFTSR. Following the standard procedures and by using the $\overline{\rm MS}$ scheme to deal with the infrared divergences, the correlator can be expressed as a expansion series over the basic vacuum condensates with increasing dimensions. Since the current quark masses of $u$ and $d$ quarks are quite small, contributions from the $u$- and $d$- quark mass terms can be safely neglected in the calculation.

Thirdly, the correlator can also be calculated by inserting a complete set of the intermediate hadronic states in the physical region. By using the conventional quark-hadron duality~\cite{Shifman:1978bx}, the hadronic expression of the correlator can be written as
\begin{align}
&{\rm Im} I_{2;\eta_q,{\rm Had}}^{(n,0)}(q^2)  = \pi \delta (q^2 - \tilde m_{\eta_q}^2)f_{\eta_q}^2\langle \xi _{2;\eta_q}^n\rangle|_\mu\langle \xi _{2;\eta_q}^0\rangle|_\mu
\nonumber\\[1ex]
& \qquad + \pi \dfrac{3}{{4{\pi ^2}(n + 1)(n + 3)}}\theta (q^2 - {s_{\eta_q}}),
\end{align}
where $\mu$ represents the initial scale at which the $\eta_q$ leading-twist LCDAs have been defined. Because of the $SU(3)_{F}$ flavour symmetry, here $\tilde m_{\eta_q}$ represents the $\eta_q$ effective mass~\cite{Bali:2014pva}, $f_{\eta_q}$ is the decay constant of $\eta_q$ and $s_{\eta_q}$ stands for the continuum threshold.

By matching the hadronic expression with the OPE results with the help of the dispersion relation, one then obtains the required sum rules. Applying the Borel transformation on both sides, we can further suppress the uncertainties caused by the unwanted contributions from both the higher-order dimensional vacuum condensates and the continuum states, and our final sum rules for the moments of $\eta_q$ leading-twist LCDA becomes
\begin{widetext}
\begin{align}
\langle \xi _{2;\eta_q}^n\rangle |_\mu \langle \xi _{2;{\eta_q}}^0 \rangle|_\mu & = \frac{M^2}{f_{\eta_q}^2} e^{\tilde m_{\eta_q}^2/M^2} \bigg\{\frac3{4\pi^2(n+1)(n+3)}(1-e^{-s_{\eta_q}/M^2}) + \frac{({m_u} + {m_d})\langle \bar qq\rangle }{M^4} + \frac{{\langle \alpha_s{G^2}\rangle }}{{12\pi {M^4}}}\frac{{1 + n\theta (n - 2)}}{{n + 1}}
\nonumber
\\
&- \frac{({m_u} + {m_d})\langle g_s\bar q\sigma TGq\rangle }{M^6}\frac{8n + 1}{18}\,+\, \frac{{\langle g_s\bar qq\rangle }^2}{M^6}\,\frac{4(2n + 1)}{18} \,-\, \frac{\langle g_s^3f{G^3}\rangle }{M^6}\,\frac{n\theta (n - 2)}{48{\pi ^2}}\,+\, \frac{{\langle g_s^2\bar qq\rangle }^2}{M^6}~\frac{2 + {\kappa ^2}}{486{\pi ^2}}
\nonumber
\\
& \times  \bigg\{ - 2\,(51n + 25)\bigg( - \ln \frac{M^2}{\mu^2}\bigg) + 3\,(17n + 35) \,+\, \theta (n - 2)\bigg[2n\bigg( - \ln \frac{M^2}{\mu^2}\bigg)+ \frac{49{n^2} + 100n + 56}{n}
\nonumber
\\
&- 25(2n + 1)\bigg[\psi \bigg(\frac{n + 1}{2}\bigg) - \psi \bigg(\frac n2\bigg)+ \ln 4\bigg]\bigg]\bigg\}\bigg\},
\label{eq:DAn}
\end{align}
\end{widetext}
where the parameter $\kappa = \langle s\bar s\rangle/\langle q\bar q\rangle$, which comes from the use of relation $g^2_s\sum\langle g_s\bar\psi\psi\rangle^2=(2+\kappa^2)\langle g^2_s\bar qq\rangle^2$ with $(\psi=u,d,s)$ in the OPE expansion. It has been shown that due to the anomalous dimension of the $n_{\rm th}$-order moment grows with the increment of $n$, contributions from the much higher moments at the large momentum transfer will be highly suppressed~\cite{Cheng:2020vwr}. Calculating the first few moments is sufficient, thus avoiding the need for further calculations. Specifically, the sum rules of the $0_{\rm th}$-order moment gives
\begin{eqnarray}
&&(\langle\xi_{2;\eta_q}^0\rangle|_\mu )^2= \frac{M^2}{f_{\eta_q}^2}e^{\tilde m_{\eta_q}^2/M^2}  \bigg\{\frac1{4 \pi^2}\bigg(1 - e^{-s_{\eta_q}/M^2}\bigg)
\nonumber\\
&&\qquad\qquad + \frac{(m_u + m_d)\langle\bar qq\rangle}{M^4} - \frac{(m_u + m_d)\langle g_s\bar q\sigma TGq\rangle }{18 M^6}
\nonumber\\
&&\qquad\qquad + \frac{\langle \alpha_s G^2 \rangle }{12\pi M^4} +\frac{4\langle g_s\bar qq\rangle^2}{18M^6} + \frac{\langle g_s^2\bar qq\rangle^2}{M^6}~\frac{2+\kappa^2}{486\pi ^2}
\nonumber\\
&&\qquad\qquad\times \bigg[-50\bigg(-\ln \frac{M^2}{\mu^2}\bigg)+105\bigg]\bigg\}.
\label{eq:DA0}
\end{eqnarray}
The effective mass ${\tilde m_{\eta_q}}$ is taken as $\sim 370~{\rm MeV}$~\cite{Bali:2014pva}. To be self-consistent, we will adopt the relation $\langle \xi _{2;\eta_q}^n\rangle |_\mu =\langle\xi_{2;{\eta_q}}^n\rangle |_\mu \langle \xi _{2;{\eta_q}}^0\rangle|_\mu/\sqrt{(\langle\xi _{2;\eta_q}^0\rangle |_\mu )^2}$ to calculate the $n_{\rm th}$-order moment~\cite{Zhong:2021epq}. The decay constant is an important input for the $B(D)\to\eta^{(\prime)}$ TFFs, which has been calculated under different methods such as the LCSR~\cite{Ball:2004ye}, the QCD sum rules (QCD SR)~\cite{DeFazio:2000my, Ali:1998eb}, the light-front quark model (LFQM)~\cite{Dhiman:2019qaa, Hwang:2010hw, Geng:2016pyr, Choi:2007se}, the lattice QCD (LQCD)~\cite{Dercks:2017lfq, Becirevic:1998ua, FermilabLattice:2014tsy, Bali:2021qem}, the Bethe-Salpeter (BS) model~\cite{Cvetic:2004qg, Wang:2005qx, Bhatnagar:2009jg}, the relativistic quark model (RQM)~\cite{Hwang:1996ha, Capstick:1989ra, Ebert:2006hj}, the non-relativistic quark model (NRQM)~\cite{Yazarloo:2016luc}, and etc.. As for the decay constant $f_{\eta_q}$, those studies show $f_{\eta_q}$ is within a broader range $[0.130,0.168]$ GeV. At present, the sum rules of the $\eta_q$ decay constant can be inversely obtained by using Eq.\eqref{eq:DA0}. The $\langle \xi _{2;{\eta_q }}^0\rangle |_\mu$ should be normalized in a suitable Borel window, which will be treated as an important criteria for determining the $\eta_q$ decay constant.

\subsection{The LCHO model for $\eta_q$ leading-twist LCDA}

The meson's LCDA can be derived from its light-cone wave-function (LCWF) by integrating its transverse components. It is helpful to first construct the $\eta_q$ leading-twist LCWF and then get its LCDA~\cite{Guo:1991eb, Huang:1994dy}. Practically, the $\eta_q$ LCWF can be constructed by using the BHL prescription~\cite{Brodsky:1981jv, Lepage:1982gd}, and the LCHO model takes the following form~\cite{Zhong:2021epq}:
\begin{align}
\psi_{2;\eta_q}(x,\mathbf k_\bot) = {\chi _{2;\eta_q }}(x,\mathbf k_\bot)\psi _{2;\eta }^R(x,\mathbf k_\bot),
\end{align}
where ${\mathbf k}_ \bot$ is the $\eta_q$ transverse momentum, ${\chi _{2;\eta_q}}(x,\mathbf k_\bot)$ stands for the spin-space WF that comes from the Wigner-Melosh rotation and the spatial WF $\psi _{2;\eta_q }^R(x,\mathbf k_\bot)$ comes from the approximate bound-state solution in the quark model for $\eta_q$. Some more explanation on the LCWF construction can be found in Ref~\cite{Zhong:2021epq}. Using the following relationship between the LCDA and LCWF,
\begin{align}
{\phi _{2;\eta_q}}(x,\mu) = \frac{2\sqrt 6 }{f_{\eta_q}}\int_0^{{|\mathbf k_\bot|}^2 \le {\mu^2}} {\frac{d^2 \mathbf k_\bot}{16\pi^3}} {\psi _{2;\eta_q}}(x,\mathbf k_\bot),     \label{Eq:phi}
\end{align}
and by integrating over the transverse momentum $\mathbf k_\bot $, one then obtains the leading-twist LCDA ${\phi _{2;\eta_q}}(x,\mu )$, i.e.
\begin{align}
&\phi _{2;\eta_q}(x,\mu)  = \frac{\sqrt 3 A_{2;\eta_q}m_q\beta _{2;\eta_q}}{2\sqrt2 \pi^{3/2} f_{\eta_q}} \sqrt{x\bar x} \varphi _{2;\eta_q}(x)
\nonumber\\
&\qquad
\times\left\{{\rm Erf} \left[\sqrt {\frac{m_q^2 + {\mu^2}}{8\beta _{2;\eta_q}^2x\bar x}}\right]- {\rm Erf} \left[\sqrt {\frac{m_q^2}{8\beta _{2;\eta_q}^2x\bar x}} \right]\right\}.    \label{eq:LCDA}
\end{align}
where $m_q=m_u=m_d$ is the constituent light-quark mass, which is around one third of the proton mass and some typical choices are $~250~{\rm MeV}$~\cite{Jaus:1991cy}, $330~{\rm GeV}$~\cite{Choi:1996mq, Ji:1992yf} and $300~{\rm MeV}$~\cite{Wu:2008yr, Wu:2011gf}, respectively.

The overall parameter $A_{2;\eta_q}$ and the transverse parameter $\beta _{2;\eta_q}$ that dominates the LCWF's transverse behavior, can be fixed according to the following two constraints. One is the normalization condition, which is the same as the pionic case, e.g.
\begin{align}
\int_0^1 {dx} \int \frac{d^2 \mathbf k_\bot}{16\pi^3} \psi_{2;\eta_q}(x,\mathbf k_\bot) = \frac{f_{\eta_q}}{2\sqrt 6}.     \label{Eq:nc}
\end{align}
Another is the probability of finding the $q\bar q$ Fock state in a meson should be not larger than $1$,
\begin{align}
P_{\eta_q} & =\int_0^1 {dx} \int \frac{d^2 \mathbf k_\bot}{16\pi^3} |\psi_{2;\eta_q}(x,\mathbf k_\bot)|^2
\nonumber\\
& = \frac{{A_{2;\eta_q}^2}m_q^2}{32\pi^2}[\varphi _{2;\eta_q}(x)]^2\Gamma \bigg[0,\frac{m_q^2}{4\beta_{2;\eta_q}^2x\bar x}\bigg].        \label{Eq:pp}
\end{align}
We adopt ${P_{\eta_q}}\approx0.3$ to carry out the following calculation, which is the same as that of pion LCWF~\cite{Huang:1994dy}. Equivalently, one can replace the constraint~\eqref{Eq:pp} by the quark transverse momentum ${\langle \mathbf k_\bot ^2\rangle _{\eta_q}}$, which is measurable and defined as~\cite{Huang:1994dy}
\begin{align}
\langle \mathbf k_\bot ^2\rangle _{\eta_q} & = \int_0^1 dx \int\frac{d^2 \mathbf k_\bot}{16\pi^3} |\mathbf k^2_\bot| \psi_{2;\eta_q}^R(x,\mathbf k_\bot)^2/P_{\eta_q}
\nonumber\\
&=\int_0^1 dx \frac{4\exp\left[- \dfrac{m_q^2}{4x\bar x\beta _{2;\eta_q}^2}\right]x\bar x \beta_{2;\eta _q}^2}{\Gamma \left[0,\dfrac{m_q^2}{4x\bar x\beta_{2;\eta_q}^2}\right]} - m_q^2
\end{align}
where the gamma function $\Gamma [s,x] = \int_0^x t^{(s-1)} e^{-t} dt$.

The function ${\varphi _{2;\eta_q}}(x)$ determines the dominant longitudinal behavior of ${\phi _{2;\eta_q}}$, which can be expanded as a Gegenbauer series as
\begin{align}
{\varphi _{2;\eta_q}}(x) =\left[1 + \sum_{n} B_n \times C_n^{3/2}(2x - 1) \right],
\end{align}
For self-consistency, the parameters $B_n$ have been observed to closely approximate their corresponding Gegenbauer moments, i.e. $B_n \sim  a_n$, especially for the first few ones~\cite{Wu:2011gf, Huang:2013yya, Wu:2012kw}. The $\eta_q$ meson Gegenbauer moments at the scale $\mu$ can be calculated by the following way
\begin{align}
a_{2;\eta_q}^n(\mu)=\frac{\int_0^1 dx{\phi _{2;\eta_q}}(x,\mu)C_n^{3/2}(2x-1)}{\int_0^1 dx6x(1-x)[C_n^{3/2}(2x-1)]^2}.
\label{eq:GM}
\end{align}
Then the Gegenbauer moments $a_{2;\eta_q}^n(\mu)$ and the moments $\langle \xi _{2;\eta_q}^n\rangle |_\mu$ satisfy the following relations,
\begin{align}
\langle \xi _{2;\eta_q}^2\rangle |_\mu&=\frac{1}{5}+\frac{12}{35}a_{2;\eta_q}^2(\mu)
\nonumber\\
\langle \xi _{2;\eta_q}^4\rangle |_\mu&=\frac{3}{35}+\frac{8}{35}a_{2;\eta_q}^2(\mu)+\frac{8}{77}a_{2;\eta_q}^4(\mu)
\nonumber\\
&\cdots .
\label{eq:xin}
\end{align}
Using the sum rules~\eqref{eq:DAn} of $\langle \xi _{2;\eta_q}^n\rangle |_\mu$, one can determine the values of $a_{2;\eta_q}^n(\mu)$, which will be used to fix $B_n$. In the following we will adopt the given two Gegenbauer moments $a^{2,4}_{2;\eta_q}$ to fix the parameters $B_{2,4}$.

\subsection{The $B(D)^+\to \eta_q \ell^+ \nu_\ell$ TFFs using the LCSR}

The LCSR approach is an effective tool in determining the non-perturbative properties of hadronic states. Here and after, we use the symbol ``$H$'' to indicate the $B(D)$-meson for convenience.

Following the LCSR approach, one should first construct a correlator with the weak current and a current with the quantum numbers of $H$ that are sandwiched between the vacuum and $\eta_q$ state. More explicitly, for $H\to \eta_q$, we need to calculate the correlator
\begin{align}
\Pi_\mu(p,q) & =i\int {d^4} x{e^{iq\cdot x}}\langle \eta_q (p)|T\{ \bar u(x){\gamma _\mu }Q(x), j_H(0)\} |0\rangle
\nonumber\\
& = \Pi[q^2, (p+q)^2] p_\mu + \tilde \Pi[q^2, (p+q)^2] q_\mu.
\label{Correlation function}
\end{align}
where the current $j_H=(m_Q \bar Q i\gamma_5 d)$ with $Q = (b,c)$-quark for $(B,D)$ meson, respectively. The LCSR calculation for the $B(D)^+ \to \eta_q$ TFFs is similar to the case of $B_s(D_s)\to \eta_s$, which has been done in Ref.\cite{Hu:2021zmy}. In the following, we will give the main procedures for self-consistency, and the interesting reader may turn to Ref.\cite{Hu:2021zmy} for more detail.

The dual property of the correlator (\ref{Correlation function}) is used to connect the two different representations in different momentum transfer regions. In the time-like region, one can insert a complete set of the intermediate hadronic states in the correlator and obtain its hadronic representation by isolating out the pole term of the lowest meson state, i.e.
\begin{align}
&\Pi_\mu^{\rm had}(p,q)=\frac{\langle\eta_q (p)|\bar u\gamma_\mu Q|H(p+q)\rangle \langle H(p+q)|\bar Qi\gamma_5q|0\rangle }{m_H^2-(p+q)^2}
\nonumber\\
&\quad +\sum\limits_{\cal H}{\frac{\langle\eta_q (p)|\bar u\gamma_\mu Q|H^{\cal H}(p+q)\rangle \langle H^{\cal H}(p+q)|\bar Q i\gamma_5q|0\rangle}{m_{H^{\cal H}}^2-(p+q)^2}}\nonumber\\
&\quad = \Pi^{\rm had}[q^2,(p+q)^2]p_\mu+\widetilde\Pi^{\rm had}[q^2,(p+q)^2]q_\mu,
\label{Eq:Hadronic Expression}
\end{align}
where the superscript ``had" and ``${\cal H}$" stand for the hadronic expression of the correlator and the continuum states of heavy meson, respectively. Here, the decay constant of $B(D)$-meson is defined via the equation, $\langle H|\bar Qi\gamma_5q|0\rangle = m_H^2 f_H/m_Q$, and by using the hadronic dispersion relations in the virtuality $(p+q)^2$ of the current in the $B(D)$ channel, we can relate the correlator to the $H\to\eta_q$ matrix element~\cite{Ball:2007hb}
\begin{align}
&\langle\eta_q (p)|\bar u\gamma_\mu Q| H(p+q)\rangle  = 2p_\mu f^{H\to\eta_q}_+(q^2)
\nonumber\\
&\qquad\qquad\qquad + q_\mu \Big( f^{H\to\eta_q}_+(q^2) + f^{H\to\eta_q}_- (q^2)\Big).
\end{align}
Due to chiral suppression, only the first term contributes to the semileptonic decay of $H\to\eta_q$ with massless leptons in the final state. Then, the hadronic expression for the invariant amplitude can be written as
\begin{align}
&\Pi[q^2,{(p+q)^2}] = \frac{2m_H^2 f_H f_+^{H\to\eta_q} (q^2)}{ [m_H^2 - (p+q)^2]}p_\mu
\nonumber\\
&\qquad\qquad\qquad
+ \int_{s_0}^\infty ds \frac{\rho^{\cal H} (q^2,s)}{s - (p+q)^2},
\end{align}
where $s_0$ is the continuum threshold parameter, $\rho^{\cal H} $ is the hadronic spectral density.

In the space-like region, the correlator can be calculated by using the operator production expansion (OPE). The OPE near the light cone $x^2 \approx 0$ leads to a convolution of perturbatively calculable hard-scattering amplitudes and universal soft LCDAs. The contributions of the three-particle part being negligible~\cite{Hu:2021zmy}, we solely focus on calculating the two-particle part here, and the corresponding matrix element is~\cite{Duplancic:2008ix}
\begin{align}
& \langle \eta_q (p)|\bar u_\alpha ^i(x)d_\beta ^j(0)|0\rangle  = \frac{i{\delta ^{ij}}}{12}{f_{\eta_q} }\int\limits_0^1 {du} {e^{iup \cdot x}}\bigg\{ {[\DS p\gamma _5]_{\beta \alpha }}\phi _{2;\eta_q}
\nonumber\\
& (u)-{[{\gamma _5}]_{\beta \alpha }}{\mu _{\eta_q} }\phi _{3;\eta_q }^p(u) + \frac{1}{6}{[{\sigma _{\nu \tau }}{\gamma _5}]_{\beta \alpha }}{p_\nu }{x_\tau }{\mu _{\eta_q} }\phi _{3;\eta_q }^\sigma (u)
\nonumber\\
& + \frac{1}{16} [\DS p\gamma _5]_{\beta \alpha }{x^2}{\phi _{4;\eta_q }}(u) - \frac{i}{2}{[\DS x\gamma _5]_{\beta \alpha }}\int\limits_0^u {\psi _{4;\eta_q }} (v)dv\bigg \}
\end{align}

The light-cone expansion for $q^2, (p+q)^2 \ll m_b^2$ (or $m_c^2$), the correlator $\Pi^{\rm OPE}$ can be written in the general form
\begin{align}
\Pi^{\rm OPE}[q^2,(p+q)^2] & = F_0(q^2,(p+q)^2)
\nonumber\\
& + \frac{\alpha_s C_F}{4\pi} F_1(q^2,(p+q)^2).
\end{align}
In the above equation, the first term is the leading-order (LO) for all the LCDAs' contributions, and the second term stands for the gluon radiative corrections to the dominant leading-twist parts.

After an analytic continuation of the light-cone expansion to physical momenta using the dispersion relation, one can equate the above two representations by the assumption of quark-hadron duality. To improve the precision of the LCSR, we also apply the Borel transformation, which results in
\begin{align}
f^{H\to \eta_q}_+ (q^2) = \frac{e^{m_H^2/M^2}}{2m_H^2 f_H} \bigg[&F_0(q^2,M^2,s_0)
\nonumber\\
&+\frac{\alpha_s C_F}{4\pi} F_1(q^2,M^2,s_0)\bigg],
\label{BM}
\end{align}
where $F_{0}$ $(F_{1})$ represents the leading-order (LO) or the next-to-leading order (NLO) contributions, respectively. Our final LCSR for the $H\to \eta_q$ TFF is
\begin{widetext}
\begin{align}
& f^{H\to\eta_q}_+(q^2) = \frac{m_Q^2 f_{\eta_q}}{2m_H^2 f_H}e^{m_H^2/M^2} \int_{u_0}^1 du e^{-s(u)/M^2}\bigg\{\frac{\phi_{2;\eta_q}(u)}u + \frac{ \mu_{\eta_q}}{m_Q}\bigg[\phi_{3;\eta_q}^p(u) + \frac16 \, \bigg(2\frac{\phi_{3;\eta_q}^\sigma (u)}u - \frac{m_Q^2+q^2-u^2m_{\eta}^2}{m_Q^2-q^2+u^2m_{\eta}^2}
\nonumber
\\
&\quad \times \frac d{du}\phi_{3;\eta_q}^\sigma (u) \!+\! \frac{4um_{\eta} ^2m_Q^2}{(m_Q^2 \!-\! q^2 \!+ u^2m_{\eta}^2)^2}\phi_{3;\eta_q}^\sigma (u)\bigg)\bigg] \!+\! ~\frac1{m_Q^2-q^2+u^2 m_{\eta}^2}\bigg[u\psi_{4;\eta_q}(u) ~+\bigg(1-
~\frac{2 u^2 m_{\eta}^2}{m_Q^2 - q^2 + u^2 m_{\eta}^2}\bigg)
\nonumber
\\
&\quad \times  \int_0^u dv \, \psi_{4;\eta_q}(v) \,- \frac{m_Q^2}{4}\frac{u}{m_Q^2 - q^2 + u^2m_{\eta}^2}\bigg(\frac{d^2}{du^2} \,- \frac{6um_{\eta} ^2}{m_Q^2-q^2+u^2m_{\eta}^2}\,\frac{d}{du} ~+ \frac{12um_{\eta}^4}{(m_Q^2 - q^2 + u^2m_{\eta} ^2)^2}\bigg)\phi_{4;\eta_q}(u)\bigg]
\nonumber
\\
&
\quad + \frac{\alpha_s C_F e^{m_H^2/M^2}}{8\pi m_H^2 f_H} F_1(q^2,M^2,s_0),
\label{Eq:fp}
\end{align}
\end{widetext}
where $\bar u=(1-u)$, $\mu_{\eta_q}=m^2_{\eta}/(m_u+m_d)$, $s(u) =  ( m_Q^2 - \bar u q^2 + u\bar u m_{\eta }^2 )/u$ and ${u_0} = \big(q^2 - {s_0} + m_{\eta } ^2 + \sqrt {(q^2 - {s_0} + m_{\eta } ^2 )^2 - 4m_{\eta } ^2(q^2 - m_Q^2)}\big)/2m_{\eta}^2$. The NLO invariant amplitude $F_1(q^2,M^2,s_0)$ can be found in Ref.\cite{Hu:2021zmy}, which is given as a factorized form of the convolutions. As will be shown below, the high-twist terms will be power suppressed and have quite small contributions to compare with those of the leading-twist terms, thus we will not discuss the uncertainties caused by the different choices of the high-twist LCDAs. For convenience, we take the $\eta_q$ twist-3 LCDAs $\phi_{3;\eta_q}^p(u)$, $\phi_{3;\eta_q}^\sigma (u)$, and the twist-4 LCDAs $\psi_{4;\eta_q}(u)$, $\phi_{4;\eta_q}(u)$, together with their parameters, as those of Ref.\cite{Duplancic:2015zna}.

Using the resultant $B(D)\to\eta^{(\prime)}$ TFFs, one can further extract the CKM matrix element $|V_{\rm cd}|$ or $|V_{\rm ub}|$ by comparing with the predictions with the experimental data, i.e. via the following equation~\cite{Fu:2013wqa}
\begin{align}
\frac{{\cal B}(H\to\eta^{(\prime)}\ell \nu_\ell )}{\tau (H)} = \int_0^{q^2_{\rm max}} {\frac{d\Gamma }{dq^2}} (H\to\eta^{(\prime)}\ell \nu_\ell),
\label{eq:CKM}
\end{align}
where $\tau (H)$ is $H$-meson lifetime, and the maximum of squared momentum transfer $q^2_{\rm max} = (m_H - m_{\eta^{(\prime)}})^2$.

\section{Numerical Analysis}\label{sec:3}

\subsection{Input parameters}

The numerical calculation is performed using the following parameters. According to the Particle Data Group (PDG)~\cite{ParticleDataGroup:2022pth}, we take the charm-quark mass $m_c({\bar m}_c)=1.27\pm0.02~{\rm GeV}$, the $b$-quark mass $m_b({\bar m}_b)=4.18^{+0.03}_{-0.02}~{\rm GeV}$; the $\eta$, $\eta'$, $D$ and $B$-meson masses are $m_\eta =0.5478~{\rm GeV}$, $m_{\eta'}=0.9578~{\rm GeV}$, $m_{D^+}=1.870~{\rm GeV}$ and $m_{B^{+}}=5.279~{\rm GeV}$, respectively; the lifetimes of $D^+$ and $B^+$ mesons are $\tau ({B^ + })=1.638\pm0.004~{\rm ps}$ and $\tau({D^ + })=1.033\pm0.005~{\rm ps}$, respectively; the current-quark-masses for the light $u$ and $d$-quarks are $m_u =2.16^{+0.49}_{-0.26} ~{\rm MeV}$ and $m_d =4.67^{+0.48}_{-0.17} ~{\rm MeV}$ at the scale $\mu =2~{\rm GeV}$. As for the decay constants $f_B$ and $f_D$, we take $f_B =0.215^{+0.007}_{-0.007}~{\rm GeV}$~\cite{Duplancic:2015zna} and $f_D=0.142\pm0.006~{\rm GeV}$~\cite{Fu:2013wqa}. The renormalization scale is set as the typical momentum flow $\mu_B=\sqrt{m^2_B-{\bar m}_b^2}\approx 3~{\rm GeV}$ for $B$-meson decay or $\mu_D \approx 1.4~{\rm GeV}$ for $D$-meson decay. We also need to know the values of the non-perturbative vacuum condensates up to dimension-six, which include the double-quark condensates $\langle q\bar q\rangle$ and $\langle g_s\bar qq\rangle ^2$, the quark-gluon condensate $\langle g_s\bar q\sigma TGq\rangle$, the four-quark condensate $\langle g_s^2\bar qq\rangle ^2 $, the double-gluon condensate $\langle \alpha_s G^2 \rangle$ and the triple-gluon condensate $\langle g_s^3fG^3\rangle$, and etc. We take their values from Refs.\cite{Colangelo:2000dp, Narison:2014wqa, Narison:2014ska},
\begin{align}
\langle q\bar q\rangle &= (-2.417_{-0.114}^{+0.227})\times  10^{-2}~{\rm GeV}^3 , \nonumber\\
\langle g_s\bar qq\rangle ^2 &= (2.082_{-0.697}^{+0.734})\times  10^{-3} ~{\rm GeV}^6 , \nonumber\\
\langle g_s\bar q\sigma TGq\rangle &=(-1.934_{-0.103}^{+0.188})\times  10^{-2}~{\rm GeV}^5 , \nonumber\\
\langle g_s^2\bar qq\rangle ^2 &= (7.420_{-2.483}^{+2.614})\times 10^{-3}~{\rm GeV}^6 , \nonumber\\
\langle \alpha_s G^2 \rangle &= 0.038\pm0.011~{\rm GeV}^4 , \nonumber\\
\langle g_s^3fG^3\rangle &\approx 0.045 ~{\rm GeV}^6 .
\end{align}
The ratio $\kappa = \langle s\bar s\rangle/\langle q\bar q\rangle= 0.74\pm0.03$ is given in Ref.~\cite{Narison:2014wqa}. When doing the numerical calculation, each vacuum condensates and current quark masses should be run from their initial values at an initial scale ($\mu_0=1$ GeV) to the required scale by applying the renormalization group equations (RGEs)~\cite{Zhong:2021epq}.

\subsection{The $\eta_q$ decay constant and the moments $\langle \xi _{2;\eta_q}^n\rangle$}

The continuum threshold parameter ($s_0$) and the Borel parameter $M^2$ are two important parameters for the sum rules analysis. When calculating the decay constant $f_{\eta_q}$, one may set its continuum threshold to be close to the squared mass of the $\eta'$ meson, i.e. $s_0=0.95\pm0.1{\rm GeV^2}$~\cite{DeFazio:2000my}. To determine the allowable $M^2$ range, e.g. the Borel window, for the $\eta_q$ decay constant, we adopt the following criteria,
\begin{itemize}
  \item The continuum contribution is less than $30\%$;
  \item The contributions of the six-dimensional condensates are no more than $5\%$;
  \item The value of $f_{\eta_q}$ is stable in the Borel window;
  \item The $\langle \xi _{2;{\eta_q}}^0\rangle |_{\mu_0}$ at the initial scale $\mu_0=1$ GeV is normalized to $1$, e.g. $\langle {\xi^0_{2;\eta_q }}\rangle|_{\mu_0}=1$.
\end{itemize}

\begin{figure}[htb]
\begin{center}
\includegraphics[width=0.45\textwidth]{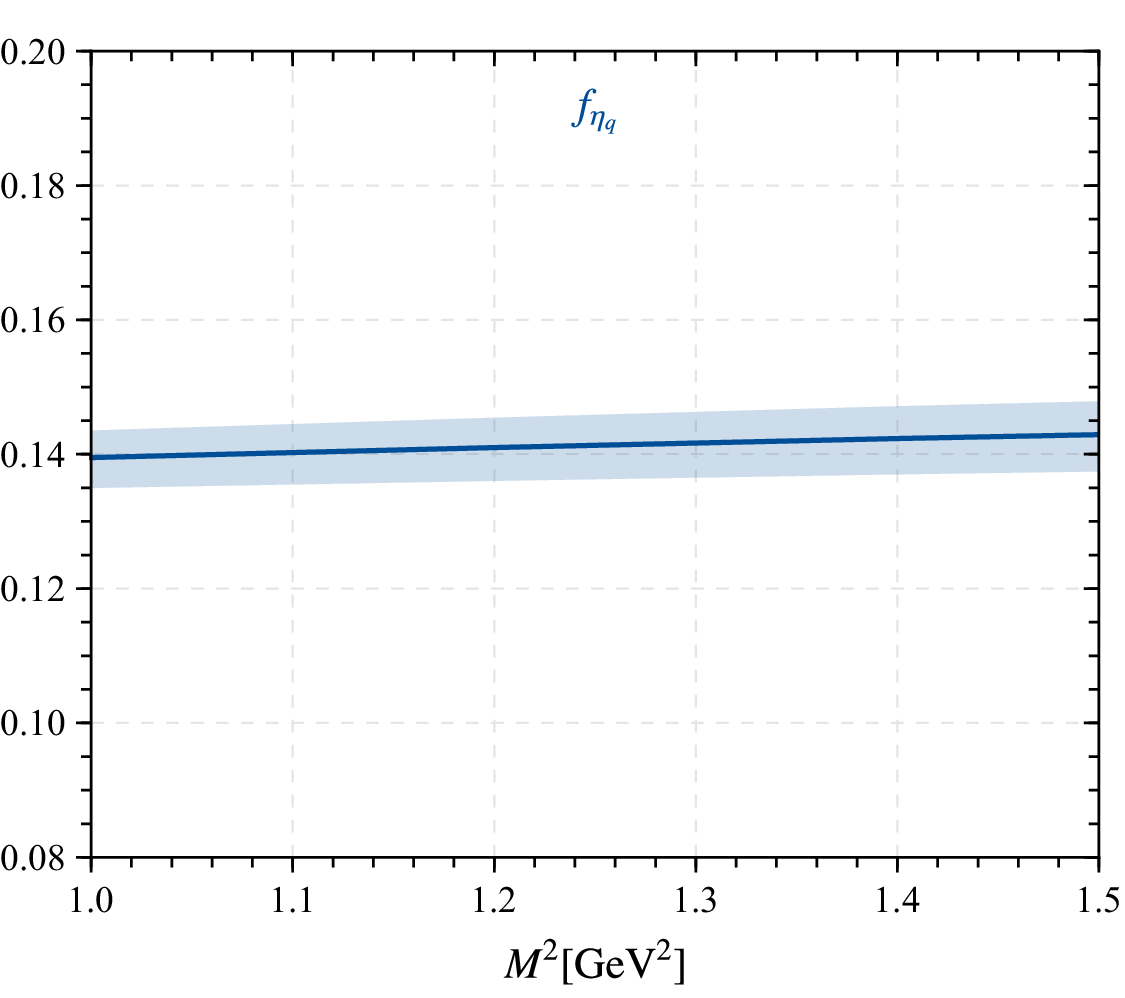}
\end{center}
\caption{(Color online) The $\eta_q$ decay constant $f_{\eta_q}$ versus the Borel parameter $M^2$, where the shaded band indicates the uncertainties from the input parameters.} \label{fig:feta}
\end{figure}

\begin{table}[htb]
\renewcommand\arraystretch{1.3}
\center
\small
\caption{The decay constant $f_{\eta_q}$ using the BFTSR approach. As a comparison, typical results derived from QCDSR and LQCD approaches have also been presented.}
\label{Tab:feta}
\begin{tabular}{l l  }
\hline
References  ~~~~~~~~ ~~~~~~~~~~~~~~~ ~~~~~~~~& $f_{\eta_q}[{\rm GeV}]$   \\
\hline
This work (BFTSR)   & $0.141_{-0.005}^{+0.005}$        \\
QCDSR~2000~\cite{DeFazio:2000my}          & $0.144_{-0.004}^{+0.004}$          \\
QCDSR~1998~\cite{Ali:1998eb}              & $0.168$                    \\
LQCD~2021~\cite{Bali:2021qem}              & $0.149_{-0.023}^{+0.034}$           \\
\hline
\end{tabular}
\end{table}

We put the decay constant $f_{\eta_q}$ versus the Borel parameter $M^2$ in Fig.~\ref{fig:feta}, where the shaded band indicates the uncertainties from the errors of all the mentioned input parameters. The decay constant is flat in the allowable Borel window, which confirms the third criterion. Using the above four criteria and the chosen continuum threshold parameter, we put the numerical results of $f_{\eta_q}$ in Table~\ref{Tab:feta}. As a comparison, we also present the predictions using the QCDSR and LQCD approaches. Our predictions are in good agreement with the QCDSR 2000~\cite{DeFazio:2000my} and the LQCD 2021 predictions within errors~\cite{Bali:2021qem}. The reason why we are slightly different from QCDSR 2000 is that their calculation only includes the contributions up to five dimensional operators, and our present one includes the dimension-6 vacuum condensation terms. Using the determined $f_{\eta_q}$, we then determine the moments of the leading-twist LCDA. Similarly, several important conditions need to be satisfied before the moments of $\eta_q$ LCDA can be determined~\cite{Hu:2021zmy}.

The Borel window is one of the important parameters to determine the moments. When determining the Borel window, it is necessary to ensure that the contributions of both the continuum state and the dimension-six condensates' contributions are sufficiently small. The lower limit of the Borel window is typically defined by the dimension-six condensates' contributions, while the upper limit is determined by the contribution of the continuum state. To find suitable Borel window for the moments, we adopt the dimension-six condensates' contributions to be no more than $5\%$ and the continuum contribution to be no more than $40\%$. More explicitly, to fix the Borel window for the first two LCDA moments $\langle \xi _{2;\eta_q}^n\rangle$ with $n=(2,4)$, we set the continuum contributions to be less than $35\%$ and $40\%$, respectively. We find that the allowable Borel windows for the two moments $\langle \xi_{2;\eta_q}^{2,4}\rangle {|_\mu }$ are $M^2\in[1.782, 2.232]$ and $M^2\in[2.740, 3.258]$, respectively. Then the first two moments $\langle \xi_{2;\eta_q}^{2,4}\rangle$ at the initial scale $\mu_{0}=1$ GeV are
\begin{align}
& \langle \xi _{2;\eta_q }^2\rangle |_{\mu_0}= 0.253\pm0.014,  \label{eq:xi0}  \\
& \langle \xi _{2;\eta_q }^4\rangle |_{\mu_0}= 0.127\pm0.010. \label{eq:xi}
\end{align}

\subsection{The LCDA $\phi_{2;\eta_q}$}

\begin{table*}[htb]
\renewcommand\arraystretch{1.3}
\center
\small
\caption{The parameters $A_{2;\eta_q}$, $\beta _{2;\eta_q}$, $B_2$, $B_4$ and the quark transverse momentum ${\langle {\mathbf k}_\bot^2\rangle _{\eta_q}}$ by taking the constituent quark mass $m_q$ to be $(250,300,350)~{\rm MeV}$, respectively.}\label{Tab:LCHO}
\begin{tabular}{l l l l l l }
\hline
~~$m_q({\rm MeV})$~~   ~~& $A_{2;\eta_q}({\rm GeV^{-1}})$~~   ~~& $\beta _{2;\eta_q}({\rm GeV})$~~
~~~~& $B_2$~~~~    ~~~~& $B_4$~~~~   ~~& ${\langle {\mathbf k}_\bot^2\rangle _{\eta_q}}({\rm GeV})$   \\   \hline
$250$    & $39.909$     & $0.589$   & $0.100$   & $0.073$   & $0.121$       \\
$300$    & $40.606$     & $0.564$   & $0.155$   & $0.108$   & $0.123$      \\
$350$    & $42.921$     & $0.549$   & $0.219$   & $0.149$   & $0.126$       \\
\hline
\end{tabular}
\end{table*}

\begin{figure}[htb]
\begin{center}
\includegraphics[width=0.45\textwidth]{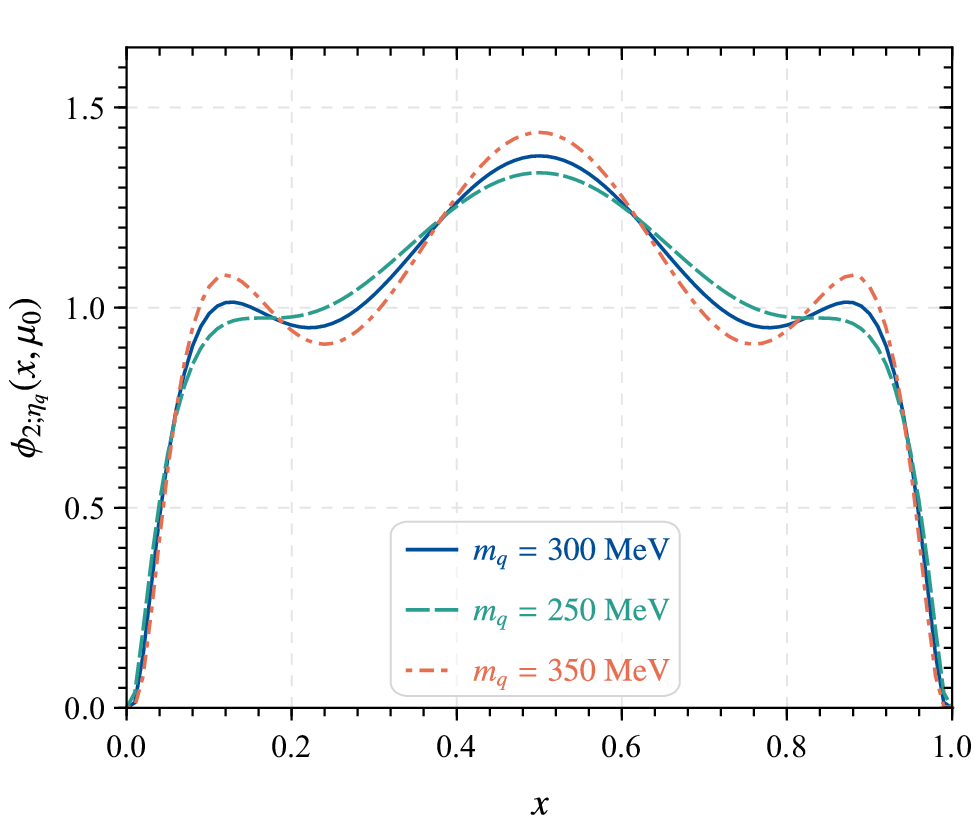}
\end{center}
\caption{(Color online) The LCHO model for the $\eta_q$ leading-twist LCDA $\phi_{2;\eta_q}$ at the scale $\mu_0=1{\rm GeV}$ with the constituent quark mass $m_q=(250, 300, 350)~{\rm MeV}$, respectively.} \label{fig:phi}
\end{figure}

\begin{figure}[htb]
\begin{center}
\includegraphics[width=0.45\textwidth]{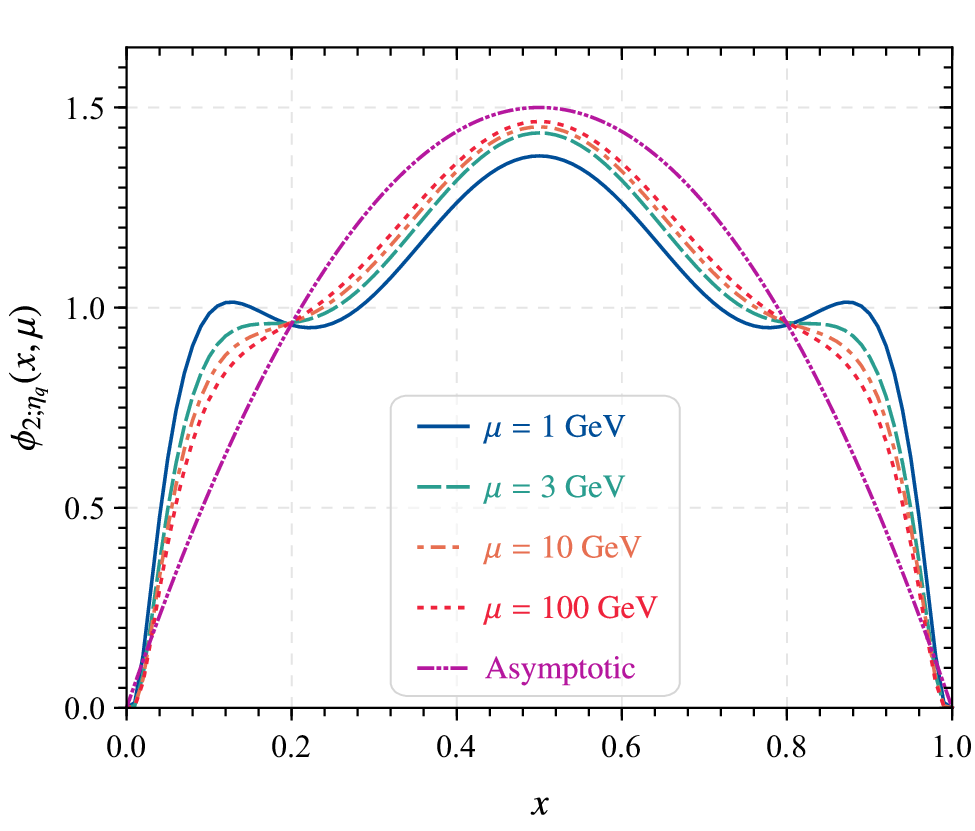}
\end{center}
\caption{(Color online) The LCHO model for the $\eta_q$ leading-twist LCDA $\phi_{2;\eta_q}(x, \mu)$ at several typical scales with $m_q=300~{\rm MeV}$.}
\label{fig:mu}
\end{figure}

Combining the normalization condition~\eqref{Eq:nc}, the probability formula for $q\bar q$ Fock state ${P_{\eta_q}}\approx0.3$, and the moments $\langle \xi _{2;\eta_q}^{(2,4)}\rangle {|_{\mu_0} }$ given in Eqs.(\ref{eq:xi0}, \ref{eq:xi}), the determined LCDA parameters are shown in Table~\ref{Tab:LCHO} and their corresponding LCDA $\phi_{2;\eta_q}$ is given in Fig.~\ref{fig:phi}. Its behavior of one peak with two humps is caused by $a_{2;\eta_q }^2({\mu _0})=0.156\pm0.042$ and $a_{2;\eta_q}^4({\mu _0})=0.055\pm0.005$, which is given by using their relations (\ref{eq:GM}) to the moments $\langle \xi _{2;\eta_q}^n\rangle$ that can be calculated by using the sum rules (\ref{eq:DAn}). In this paper, we take $m_q=300~{\rm MeV}$ to do the following calculation and use $\Delta m_q=\pm 50$ MeV to estimate its uncertainty. Table~\ref{Tab:LCHO} shows that the parameters $B_2$ and $B_4$ and the quark transverse momentum ${\langle {\mathbf k}_\bot^2\rangle _{\eta_q}}$ increase with the increment of constituent quark mass, but the harmonious parameter $\beta _{2;\eta_q}$ decreases gradually. Experimentally the average quark transverse momentum of pion, ${\langle {\mathbf k}_ \bot^2\rangle_{\pi}}$, is of the order $(300~{\rm MeV})^2$~\cite{Metcalf:1979iw}. It is reasonable to require that $\sqrt{\langle {\mathbf k}_\bot^2\rangle_{\eta_q}}$ have the value of about a few hundreds ${\rm MeV}$~\cite{Huang:1994dy}. For the case of $m_q=300\pm50~{\rm MeV}$, we numerically obtain ${\langle {\mathbf k}_\bot^2\rangle_{\eta_q}} \approx (351^{+4}_{-3}~{\rm MeV})^2$, which is reasonable and in some sense indicates the inner consistency of all the LCHO model parameters. Moreover, by using the RGE, one can get the $\phi_{2;\eta_q}(x, \mu)$ at any scale $\mu$~\cite{Zhong:2021epq}. Fig.~\ref{fig:mu} shows the LCDA $\phi_{2;\eta_q}$ at several typical scales with $m_q=300~{\rm MeV}$. At low scale, it shows double humped behavior and when the scale $\mu$ increases, the shape of $\phi_{2;\eta_q}$ becomes narrower; and when $\mu\to\infty$, it will tends to single-peak asymptotic behavior for the light mesons, $\phi^{\rm as}_{\eta_q}(x,\mu)|_{\mu\to\infty}=6x(1-x)$.

\begin{figure}[htb]
\begin{center}
\includegraphics[width=0.45\textwidth]{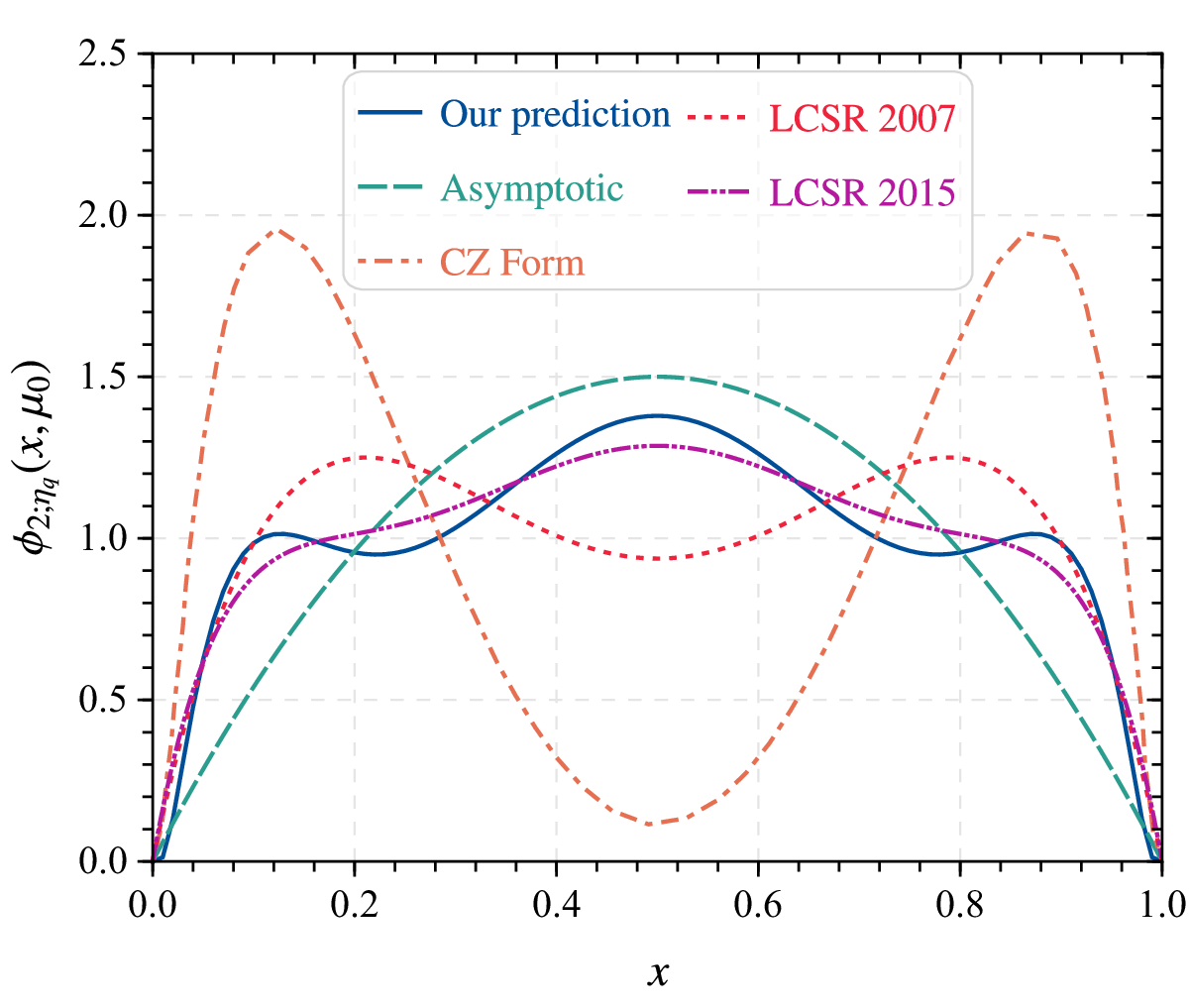}
\end{center}
\caption{(Color online) The $\eta_q$ meson leading-twist LCDA $\phi_{2;\eta_q}(x,\mu_0)$. As a comparison, the asymptotic and CZ forms~\cite{Lepage:1980fj, Chernyak:1981zz} and the one derived using the LCSR approach~\cite{Ball:2007hb, Duplancic:2015zna} are also presented. }
\label{fig:phi1}
\end{figure}

We make a comparison of the properties of the LCHO model of the leading-twist LCDA $\phi_{2;\eta_q}$ with other theoretical predictions in Fig.~\ref{fig:phi1}. Fig.~\ref{fig:phi1} gives the results for $\mu=\mu_0=1$ GeV, where the asymptotic form~\cite{Lepage:1980fj}, the CZ form~\cite{Chernyak:1981zz} and the behaviors given by the LCSR 2007~\cite{Ball:2007hb} and LCSR 2015~\cite{Duplancic:2015zna} are presented. For the LCSR 2007 result, its double peaked behavior is caused by the keeping its Gegenbauer expansion only with the first term together with the approximation $a_{2;\eta_q}^2({\mu _0})=a_{2;\eta'_q}^2({\mu _0})=0.25$~\cite{Ball:2007hb}. For the LCDA used in LCSR 2015~\cite{Duplancic:2015zna}, its behavior is close to our present one. It is obtained by using the approximation that the leading-twist LCDA $\phi_{2;\eta_q}$ has the same behavior as that of the pion leading-twist LCDA $\phi_{2;\pi}$, e.g. $a_{2;\eta_q}^2({\mu _0})=a_{2;\pi}^2({\mu _0})=0.17$ and $a_{2;\eta_q}^4({\mu _0})=a_{2;\pi}^4({\mu _0})=0.06$, which are consistent with our Gegenbauer moments within errors~\footnote{Since the leading-twist parts dominant the TFFs, this consistency also explains why our following LCSR predictions for the TFFs are close in shape with those of Ref.\cite{Duplancic:2015zna}.}.

\subsection{The TFFs and observable for the semileptonic decay ${B(D)}^+\to\eta^{(\prime)}\ell^+\nu_\ell$}

One of the most important applications of the $\eta_q$-meson LCDAs is the semileptonic decay $H^+\to\eta^{(\prime)}\ell^+\nu_\ell$, whose main contribution in the QF scheme comes from the $|\eta_q\rangle$-component. Here $H^+$ stands for $B^+$ or $D^+$, respectively. And to derive the required $H^+\to\eta^{(\prime)}$ TFFs, we take the mixing angle $\phi=(41.2^{+0.05}_{-0.06})^\circ$~\cite{Hu:2021zmy}.

\begin{figure*}[htb!]
\begin{center}
\includegraphics[width=0.243\textwidth]{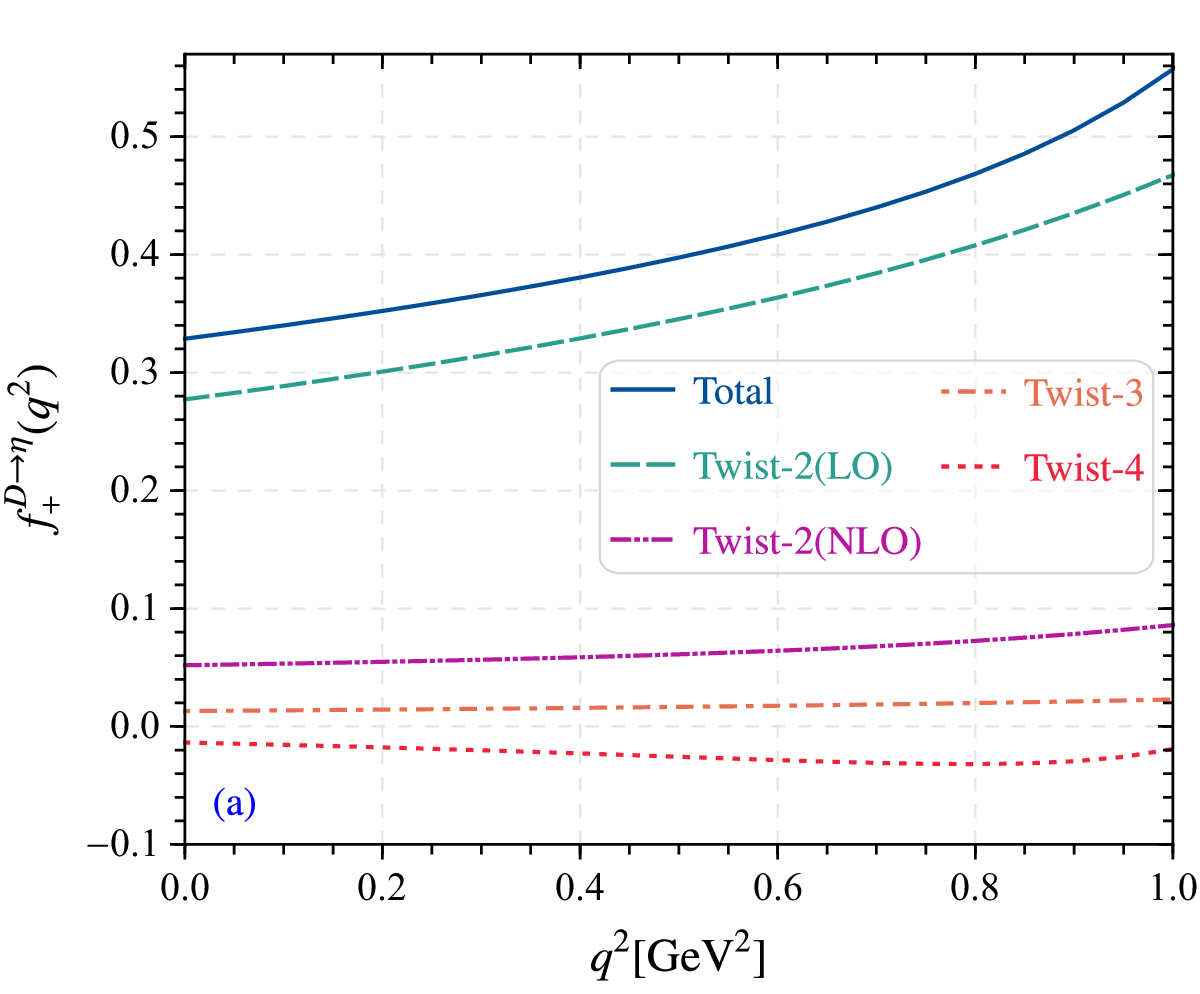}
\includegraphics[width=0.243\textwidth]{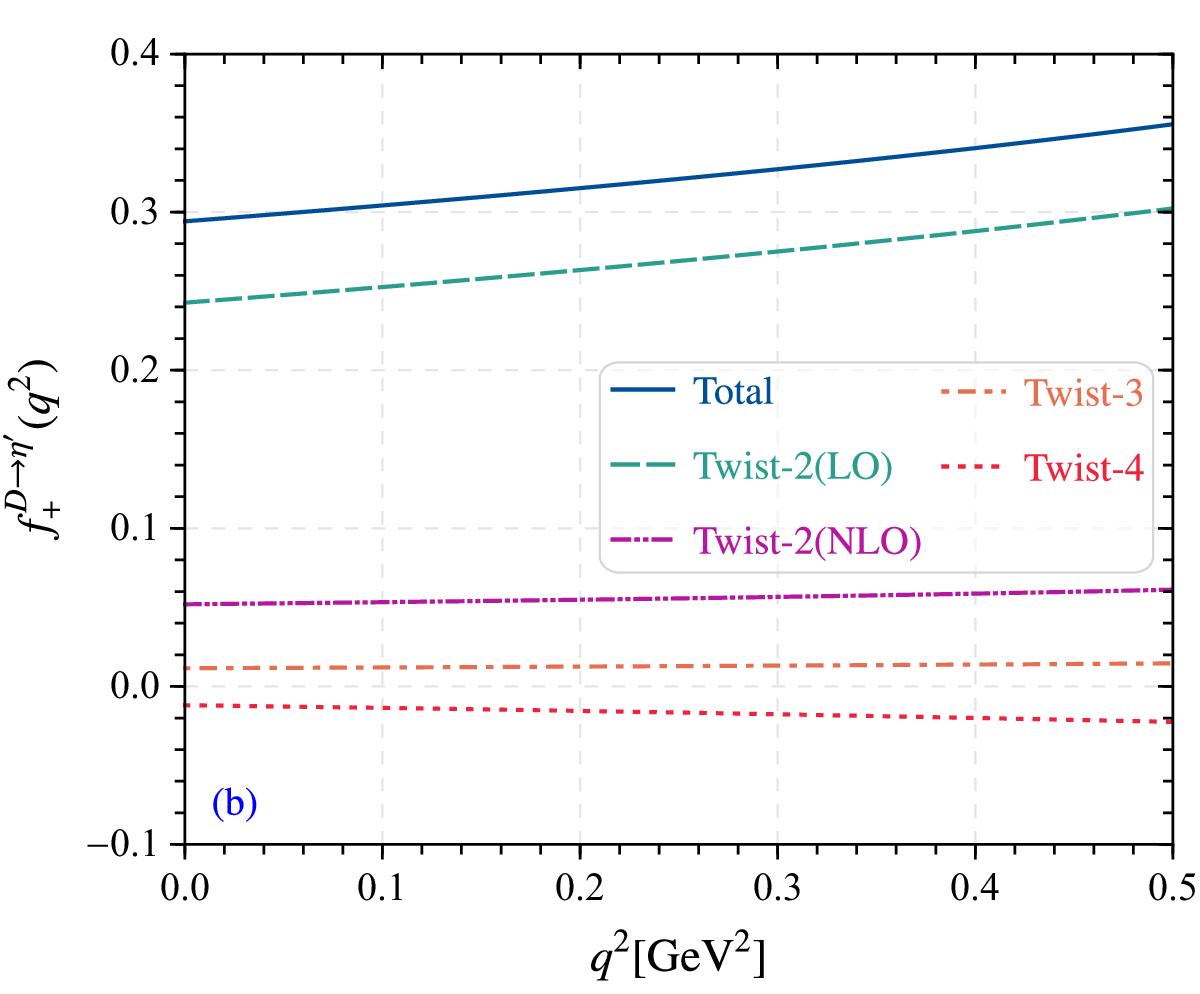}
\includegraphics[width=0.24\textwidth]{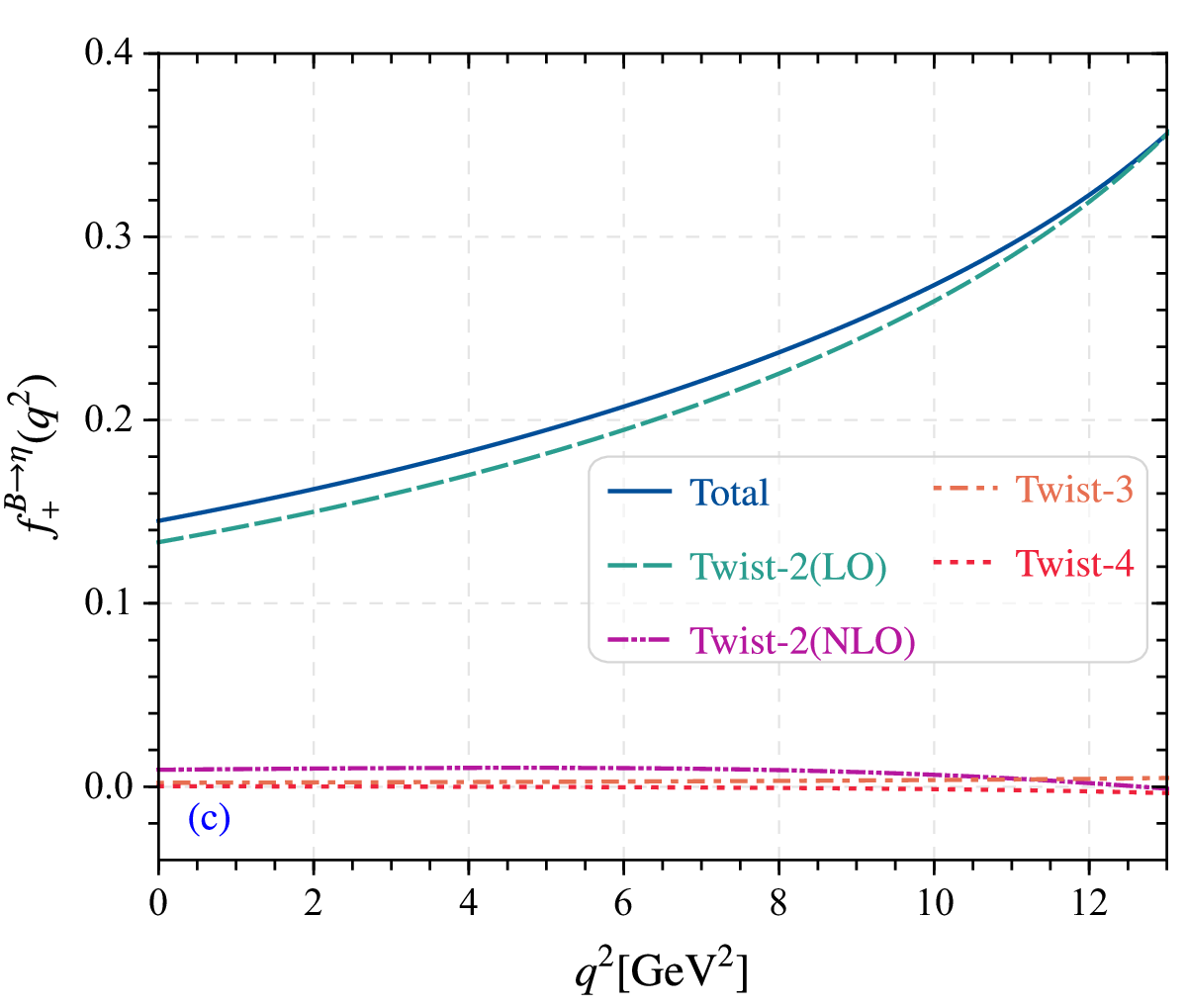}
\includegraphics[width=0.24\textwidth]{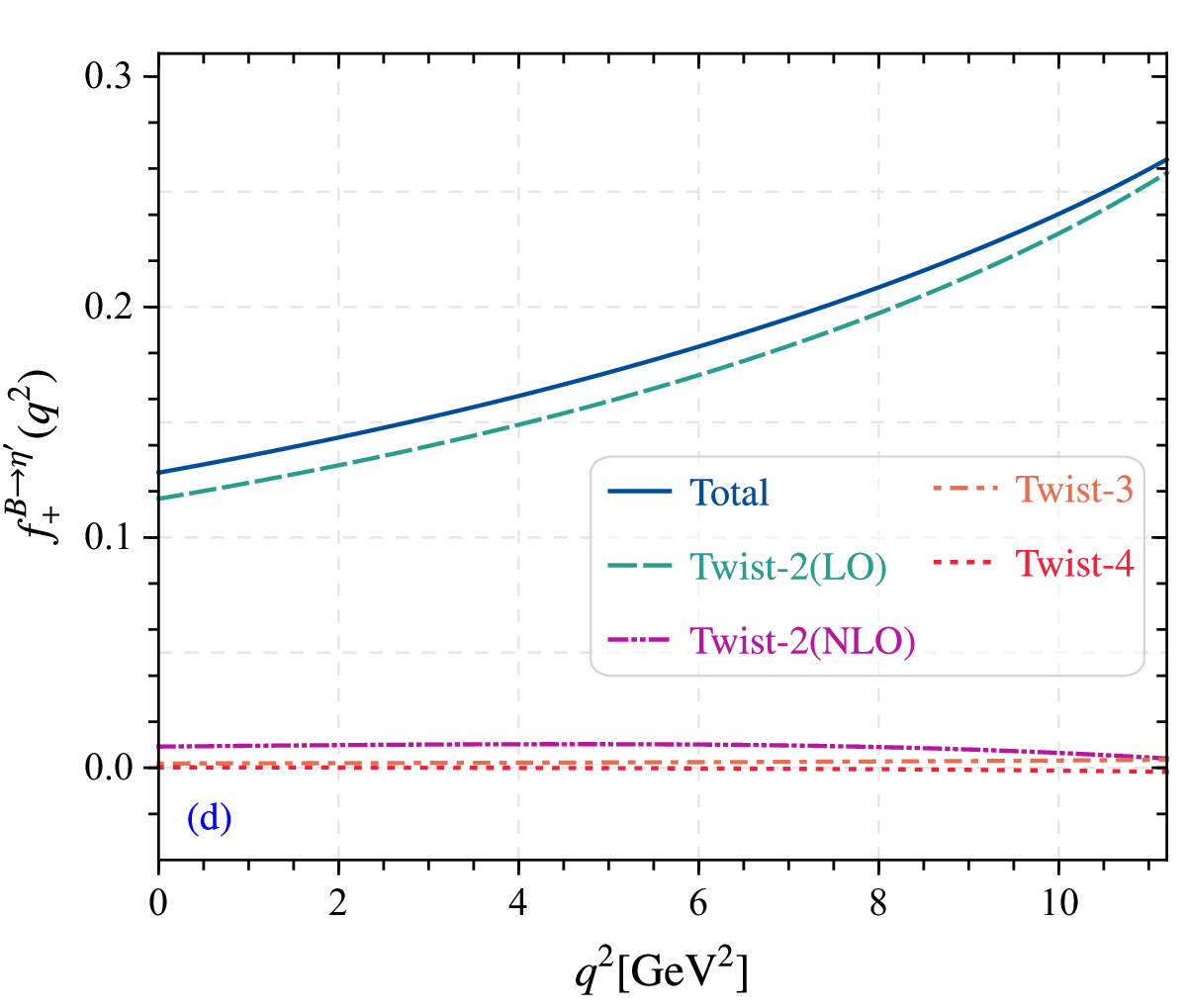}
\end{center}
\caption{(Color online) LCSR predictions on the TFFs $f_+^{H\to\eta^{(\prime)}}(q^2)$ with $H=B^+$ or $D^+$ in the allowable $q^2$ range, where the contributions from the twist-2, twist-3, twist-4 LCDAs are given separately. The twist-2 terms are given up to NLO QCD corrections. }
\label{fig:tw}
\end{figure*}

The continuum threshold $s^{H\to\eta^{(\prime)}}_0$ and Borel parameters $M^2$ are two important parameters for the LCSR of the TFFs. As usual choice of treating the heavy-to-light TFFs, we set the continuum threshold as the one near the squared mass of the first excited state of $D$ or $B$-meson, accordingly. And to fix the Borel window for the TFFs, we require the contribution of the continuum states to be less than $30\%$. The determined values agree with Refs.\cite{Duplancic:2015zna, Wang:2014vra}, and we will take the following values to do our discussion
\begin{align}
& s_0^{D\to\eta}=   7.0\pm0.5~{\rm GeV^2},   && M^2_{D\to\eta}  = 3.0\pm0.5~{\rm GeV}.
\nonumber\\
& s_0^{D\to\eta'} = 7.0\pm0.5~{\rm GeV^2},   && M^2_{D\to\eta'} = 3.0\pm0.5~{\rm GeV}.
\nonumber\\
& s_0^{B\to\eta} =  37.0\pm 1.0~{\rm GeV^2},   && M^2_{B\to\eta}  = 18.0\pm2.0~{\rm GeV}.
\nonumber\\
& s_0^{B\to\eta'} = 37.0\pm1.0~{\rm GeV^2},  && M^2_{B\to\eta'} = 18.0\pm2.0~{\rm GeV}. \nonumber
\end{align}

\begin{table}[htb]
\renewcommand\arraystretch{1.3}
\center
\small
\caption{Typical theoretical predictions on the TFFs $f_+^{H\to\eta^{(\prime)}}(0)$ at the large recoil point $q^2=0$. } \label{Tab:fq}
\begin{tabular}{l l l }
\hline
~~~~~~~~~~~~~~~~~~~~~~~~~~~~  & $f_+^{B\to\eta } (0)$~~~~~~~~~~~~~~~~  & $f_+^{B\to\eta' }(0)$   \\      \hline
This work (LCSR)   & $0.145_{-0.010}^{+0.009}$     & $0.128_{-0.009}^{+0.008}$   \\
LCSR 2007~\cite{Ball:2007hb}         & $0.229\pm 0.035$    & $0.188\pm 0.028$ \\
LCSR 2015~\cite{Duplancic:2015zna}         & $0.168_{-0.047}^{+0.041}$    & $0.130_{-0.032}^{+0.036}$  \\
pQCD~\cite{Charng:2006zj}          & $0.147$       & $0.121$   \\
CLF~\cite{Chen:2009qk}             & $0.220\pm0.018$   & $0.180\pm0.016$   \\
LCSR 2013~\cite{Offen:2013nma}         & $0.238\pm 0.046$             & $0.198\pm 0.039$   \\  \hline
                      & $f_ +^{D\to\eta } (0)$& $f_+ ^{D\to \eta' }(0)$   \\   \hline
This work (LCSR)   & $0.329_{-0.015}^{+0.021}$     & $0.294_{-0.015}^{+0.021}$   \\
LCSR 2015~\cite{Duplancic:2015zna}       & $0.429_{-0.141}^{+0.165}$    & $0.292_{-0.104}^{+0.113}$ \\
BES-III 2020~\cite{Ablikim:2020hsc}         & $0.39\pm0.04\pm0.01$     & -\\
LFQM~\cite{Verma:2011yw}              & $0.39$            & $0.32$   \\
CCQM~\cite{Ivanov:2019nqd}               & $0.36(5)$       & $0.36(5)$   \\
LCSR 2013~\cite{Offen:2013nma}           & $0.552(51)$    & $0.458(105)$ \\  \hline
\end{tabular}
\end{table}

Using Eqs.(\ref{Eq:ma0}, \ref{Eq:ma}) together with the LCSR (\ref{Eq:fp}) for the TFF $f^{H\to\eta_q}_+(q^2)$, we then get the results for $f_+^{H\to\eta^{(\prime)}}(q^2)$, where $H$ represents $B$ or $D$, respectively. Fig.~\ref{fig:tw} shows how the total TFFs $f_+^{H\to\eta^{(\prime)}}(q^2)$ change with the increment of $q^2$, in which the twist-$2$ up to NLO QCD corrections, the twist-$3$ and the twist-$4$ contributions have been presented separately. The non-local operator matrix elements in LCSR can factorization into the universal hadron distribution amplitude, the latter term is suppressed by powers of a small parameter $1/M^2$ as compared with the previous term, thereby suppressing the contribution from higher-order twist. Fig.~\ref{fig:tw} shows that the twist-$2$ terms dominant the TFFs. We also find that the NLO QCD corrections to the twist-$2$ terms are sizable and should be taken into consideration for a sound prediction. For examples, at the large recoil point, the twist-$2$ NLO terms give about $15.8\%$ $(17.6\%)$ and $6.4\%$ $(7.2\%)$ contributions to the total TFFs $f_+^{D\to\eta^{(\prime)}}(0)$ and $f_+^{B\to\eta^{(\prime)}}(0)$, respectively. Table~\ref{Tab:fq} gives our present LCSR predictions for the TFFs $f_+^{D\to\eta^{(\prime)}}(0)$ and $f_+^{B\to\eta^{(\prime)}}(0)$. As a comparison, we have also presented the results derived from various theoretical approaches and experimental data in Table~\ref{Tab:fq}, including the LCSR approach~\cite{Ball:2007hb, Duplancic:2015zna, Offen:2013nma}, the pQCD approach~\cite{Charng:2006zj}, the covariant light front (CLF) approach~\cite{Chen:2009qk}, the light front quark model (LFQM) approach~\cite{Verma:2011yw}, the covariant confining quark mode (CCQM) approach~\cite{Ivanov:2019nqd}, and the BES-III Collaboration~\cite{Ablikim:2020hsc}. The uncertainties of the TFFs $f_+^{H\to\eta^{(\prime)}}(0)$ caused by different input parameters are listed as follows,
\begin{eqnarray}
f_+^{B\to\eta}(0)&=& 0.145(_{-0.004}^{+0.004})_{s_0}(_{-0.002}^{+0.002})_{M^2}(_{-0.007}^{+0.007})_{m_b f_B}
\nonumber \\
&& (_{-0.005}^{+0.005})_{f_{\eta_q}}(_{-0.0001}^{+0.0001})_{\phi}
\nonumber \\
&=& 0.145_{-0.010}^{+0.009},
\\
\nonumber \\
f_+^{B\to\eta'}(0)&=& 0.128(_{-0.003}^{+0.003})_{s_0}(_{-0.002}^{+0.002})_{M^2}(_{-0.006}^{+0.006})_{m_b f_B}
\nonumber \\
&& (_{-0.005}^{+0.005})_{f_{\eta_q}}(_{-0.0001}^{+0.0002})_{\phi}
\nonumber \\
&=& 0.128_{-0.009}^{+0.008},
\\
\nonumber \\
f_+^{D\to\eta}(0)&=& 0.329 (_{-0.004}^{+0.003})_{s_0} (_{-0.005}^{+0.009})_{M^2} (_{-0.009}^{+0.016})_{m_c f_D}
\nonumber \\
&& (_{-0.010}^{+0.010})_{f_{\eta_q}}(_{-0.0003}^{+0.0002})_{\phi}
\nonumber \\
&=& 0.329_{-0.015}^{+0.021},
\\
\nonumber \\
f_+^{D\to\eta'}(0)&=& 0.294(_{-0.004}^{+0.003})_{s_0}(_{-0.005}^{+0.009})_{M^2}(_{-0.011}^{+0.017})_{m_c f_D}
\nonumber \\
&& (_{-0.009}^{+0.009})_{f_{\eta_q}}(_{-0.0003}^{+0.0002})_{\phi}
\nonumber \\
&=& 0.294_{-0.015}^{+0.021}.
\end{eqnarray}
Here the second equations show the squared averages of the errors from all the mentioned error sources.
The errors are mainly caused by $f_{\eta_q}$ and $m_Q f_H$, and error caused by the mixing angle $\phi$ is quite small. As a common limitation of phenomenological quark models, accurately quantifying the theoretical uncertainty of predictions in LCSR is challenging. The net errors of these parameters are about $10\%-20\%$.

\begin{table}[htb]
\renewcommand\arraystretch{1.3}
\center
\small
\caption{Fitting parameters $b_1$ and $b_2$ for the TFFs $f_+^{H\to\eta^{(\prime)}}(q^2)$, where all input parameters are set to be their central values. $\Delta$ is the measure of the quality of extrapolation.}
\label{Tab:fit}
\begin{tabular}{c c c c c }
\hline
&~~ $f_+^{D\to\eta}(q^2)$  ~&~ $f_+^{D\to\eta'}(q^2)$   ~&~ $f_+^{B\to\eta}(q^2)$   ~&~ $f_+^{B\to\eta'}(q^2)$
\\   \hline
$b_1$      & $-0.033$     & $-0.680$         & $-0.392$         & $-0.397$   \\
$b_2$      & $37.901$     & $23.961$         & $-0.108$         & $-0.308$   \\
$\Delta$   & $0.761\%$    & $0.026\%$        & $0.341\%$        & $0.062\%$   \\   \hline
\end{tabular}
\end{table}

\begin{figure*}[htb]
\begin{center}
\includegraphics[width=0.24\textwidth]{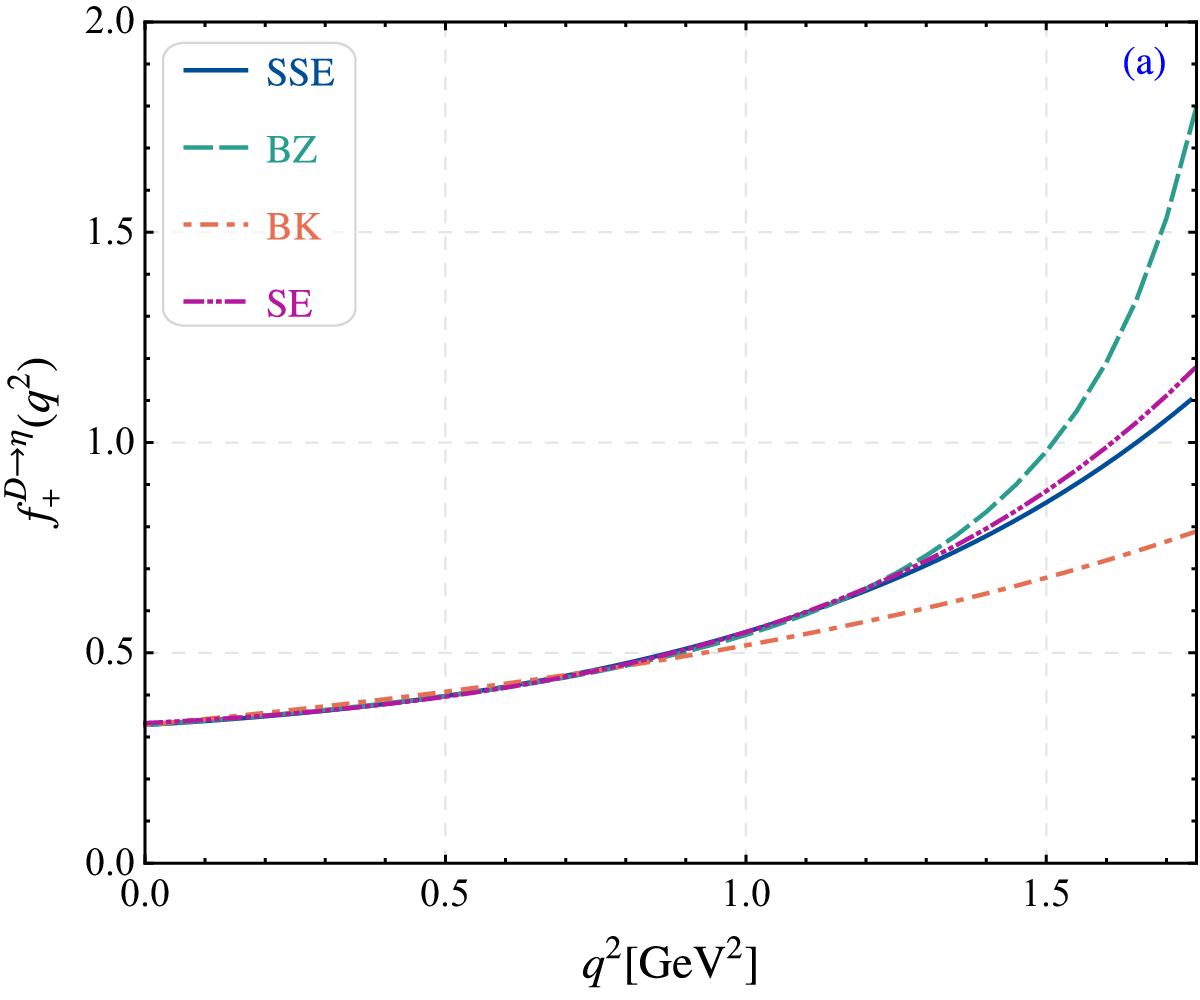}
\includegraphics[width=0.24\textwidth]{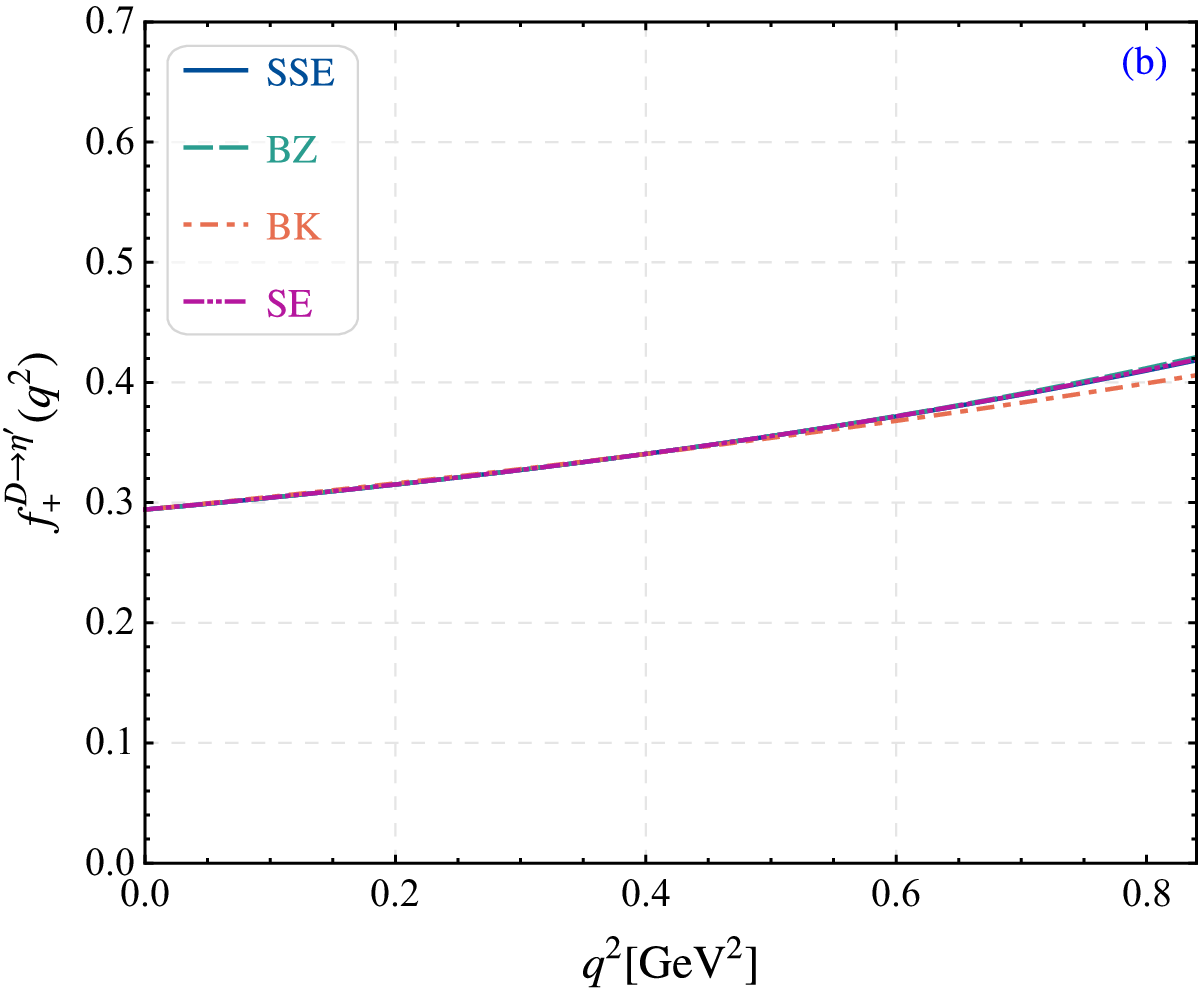}
\includegraphics[width=0.24\textwidth]{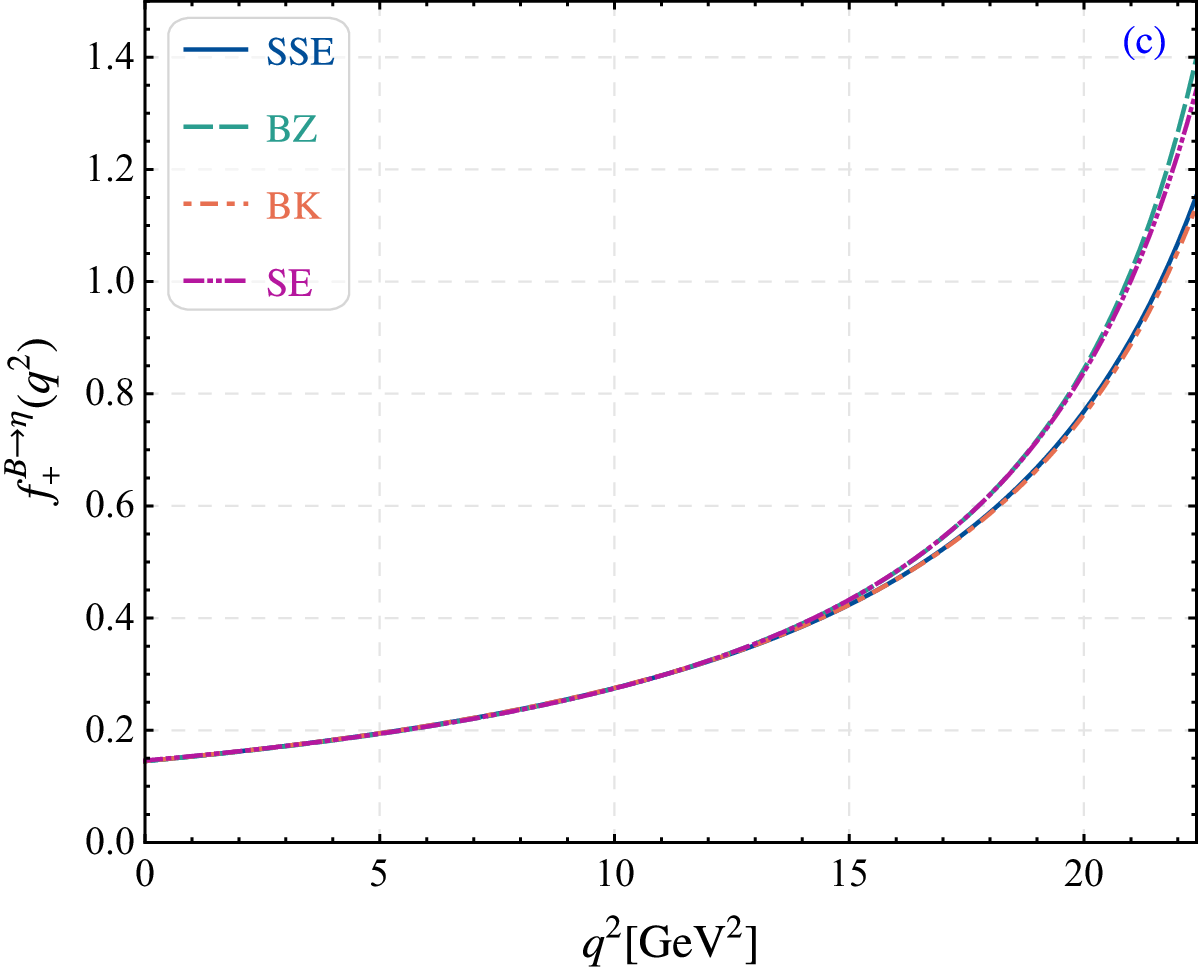}
\includegraphics[width=0.24\textwidth]{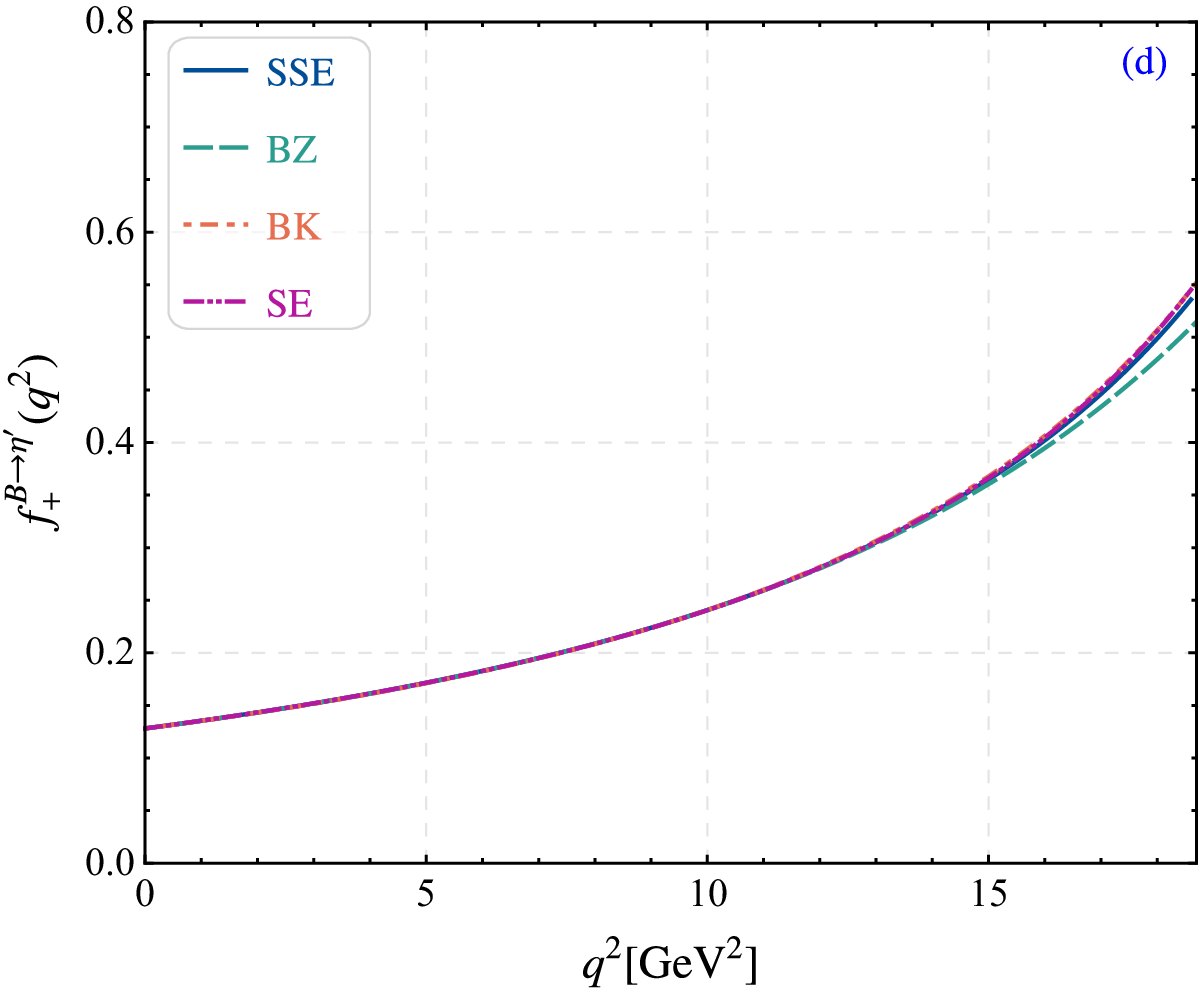}
\end{center}
\caption{(Color online) The central value of the TFFs $f_+^{H\to\eta^{(\prime)}}(q^2)$ is obtained using different parameterizations throughout the entire $q^2$-region.}
\label{fig:fit}
\end{figure*}
\begin{figure*}[htb]
\begin{center}
\includegraphics[width=0.24\textwidth]{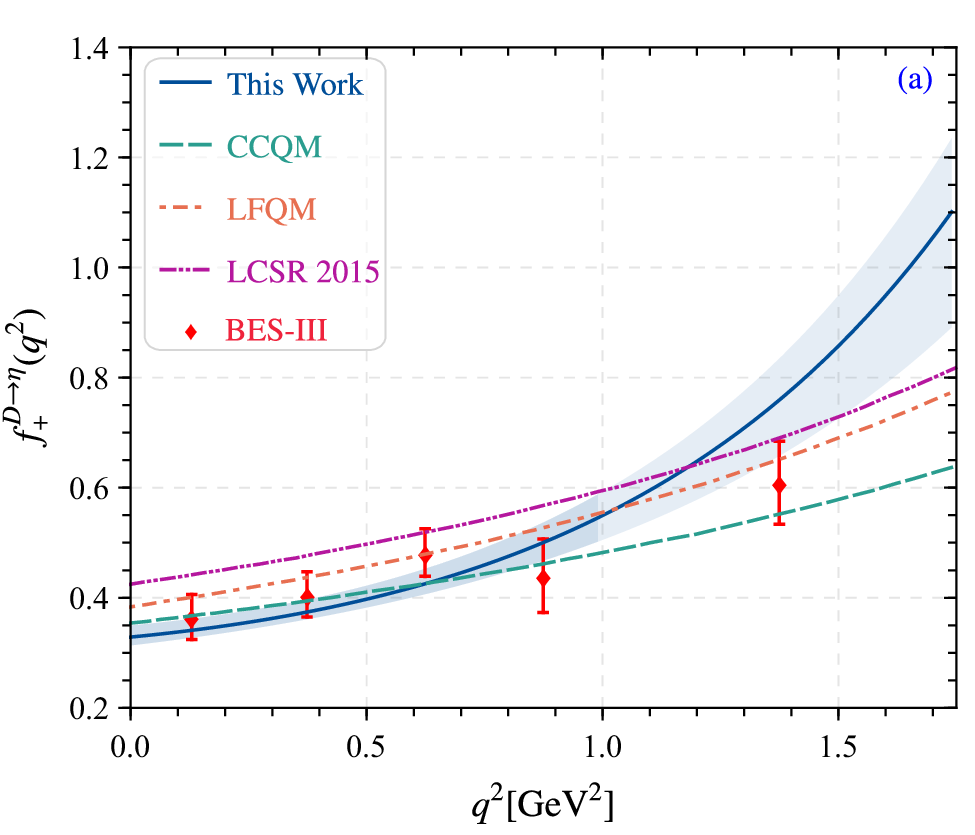}
\includegraphics[width=0.24\textwidth]{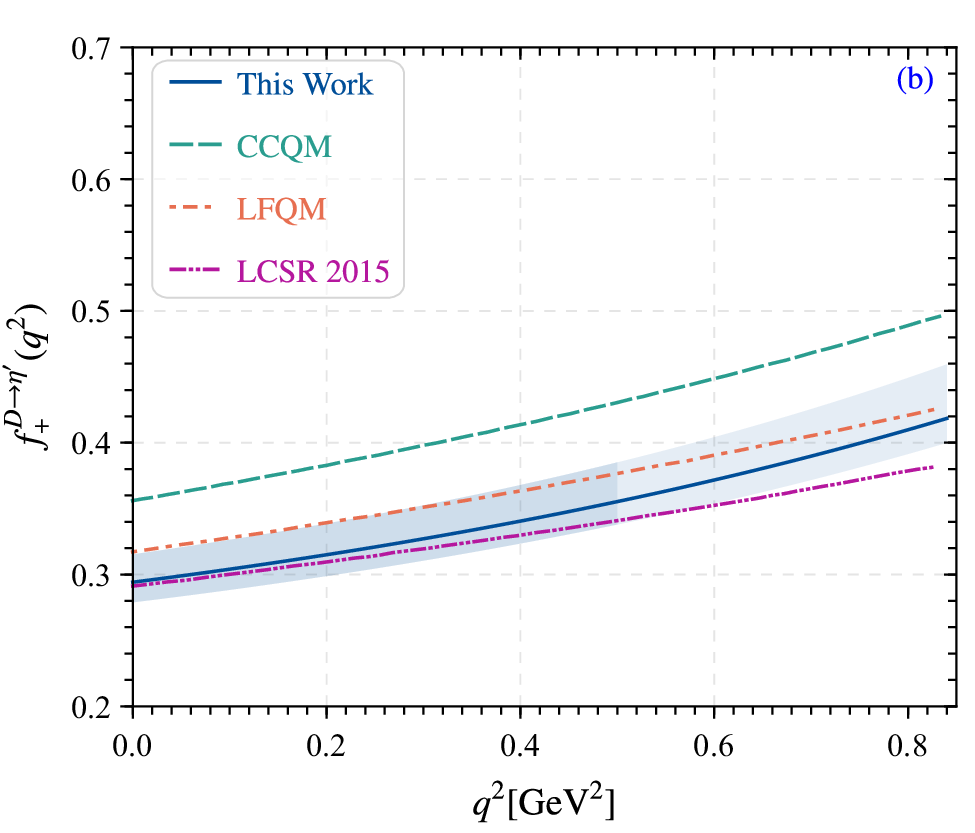}
\includegraphics[width=0.24\textwidth]{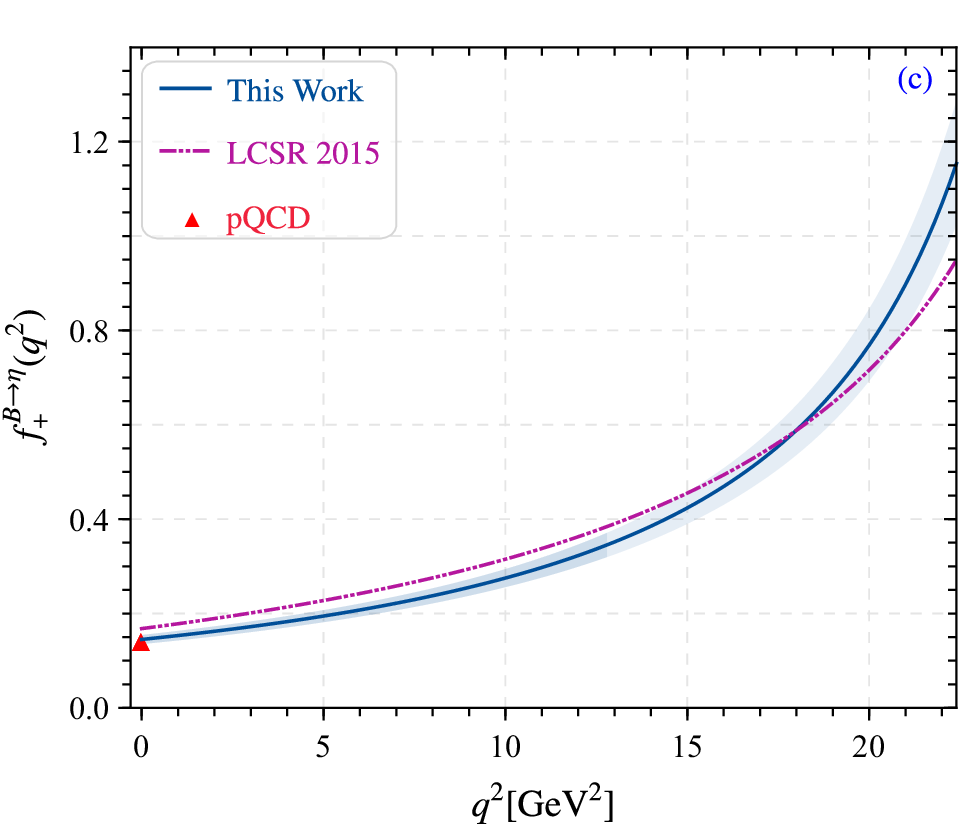}
\includegraphics[width=0.24\textwidth]{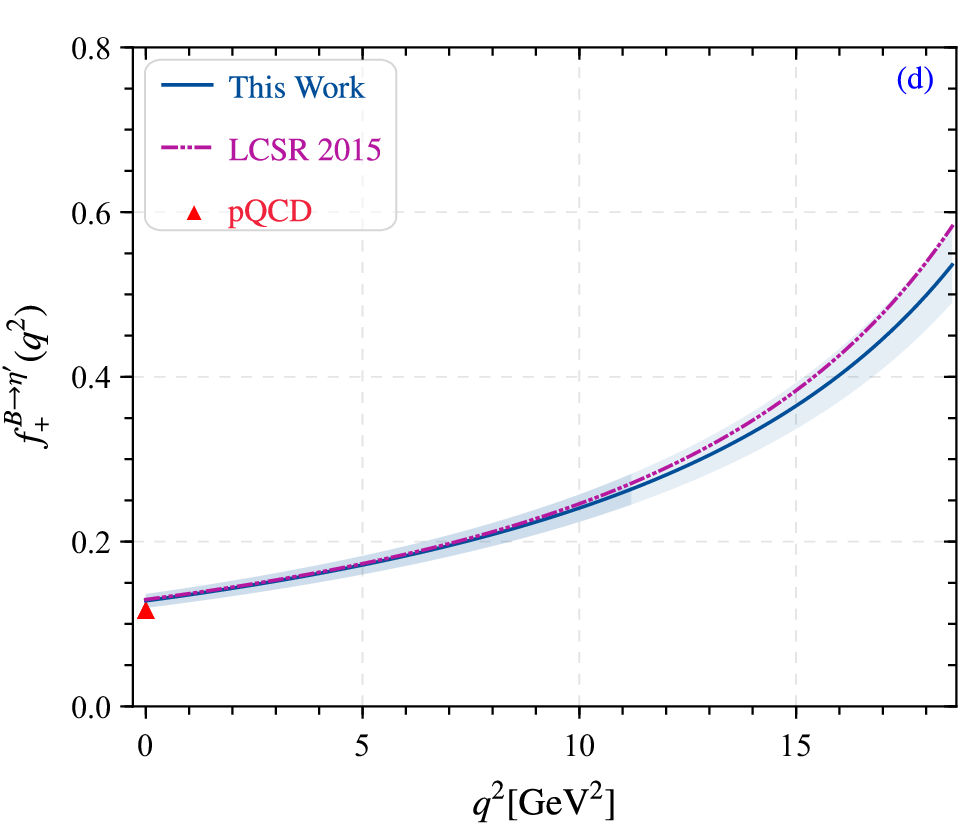}
\end{center}
\caption{(Color online) The TFFs $f_+^{H\to\eta^{(\prime)}}(q^2)$ in whole $q^2$-region, where the solid line is the central value and the shaded band shows its uncertainty. The darker part of the shaded band is the LCSR prediction, and the remaining part is the extrapolated result. As a comparison, predictions using different theoretical approaches and the experimental data, such as CCQM~\cite{Ivanov:2019nqd}, LFQM~\cite{Verma:2011yw}, LCSR~\cite{Duplancic:2015zna}, pQCD~\cite{Charng:2006zj} and BESIII collaboration~\cite{Ablikim:2020hsc}, are also presented.}
\label{fig:fq}
\end{figure*}

The physically allowable ranges of the above four heavy-to-light TFFs are $m_\ell ^2 \le {q^2} \le {({m_{D^ + }} - {m_\eta })^2} \approx 1.75~{\rm GeV^2}$, $m_\ell^2\simeq 0 \le {q^2} \le {({m_{D^ + }} - {m_{\eta' }})^2} \approx 0.84~{\rm GeV^2}$, $m_\ell ^2 \le {q^2} \le {({m_{B^ + }} - {m_\eta })^2} \approx 22.40~{\rm GeV^2}$ and $m_\ell ^2 \le {q^2} \le {({m_{B^ + }} - {m_{\eta' }})^2} \approx 18.67~{\rm GeV^2}$, respectively. For light leptons, we have $m_\ell^2\simeq 0$. The LCSR approach is applicable in low and intermediate $q^2$ region, which however can be extended to whole $q^2$ region via proper extrapolation approaches. In the present paper, we adopt the converging simplified series expansion (SSE) proposed in Refs.\cite{Bourrely:2008za, Bharucha:2010im} to do the extrapolation, which suggest a simple parameterization for the heavy-to-light TFF, e.g.
\begin{align}
f_+^{H\to \eta^{(\prime)}}(q^2) = \frac1{1 - q^2/m_{R^*}^2}\sum\limits_k b_k z^k(t,t_0)
\end{align}
where $m_{R^*}=m_{B^*}=5.325{\rm GeV}$ $(m_{D^*}=2.010{\rm GeV})$~\cite{Duplancic:2015zna} are vector meson resonances, $z(t,t_0)$ is a function
\begin{align}
z(t,t_0) = \frac{\sqrt {t_+ - t} - \sqrt{t_+ - t_0} }{\sqrt {t_+ -t} + \sqrt{t_+  - t_0} }.
\end{align}
Here $t_\pm = (m_{H^+} \pm {m_{\eta^{(\prime)}}})^2$ and $t_0 = t_+ (1 - \sqrt {1 - t_-/t_+})$ is a free parameter. The free parameter $b_k$ can be fixed by requiring $\Delta <1\%$, where the parameter $\Delta$ is used to measure the quality of extrapolation and it is defined as
\begin{align}
\Delta  = \frac{\sum\nolimits_t |F_i(t) - F_i^{\rm fit}(t)|}{\sum\nolimits_t |F_i(t)|} \times 100,
\end{align}
where $t \in [0,\frac{1}{40}, \cdots ,\frac{40}{40}] \times 13.0(1.0)~{\rm GeV}$ for the case of $\eta$-meson, $t \in [0,\frac{1}{40}, \cdots ,\frac{40}{40}] \times 11.2(0.5)~{\rm GeV}$ for the case of $\eta'$-meson. The two coefficients $b_{1,2}$ with all input parameters are set as their central values are listed in Table~\ref{Tab:fit}. The qualities of extrapolation parameter $\Delta$ are less than $\sim 0.8\%$. It is noted that the fitting parameters $b_1$ and $b_2$ were obtained by rigorous fitting LCSR data. In the decay of $B$-meson, the difference between $b_1$ and $b_2$ is small, whereas in the decay of $D$-meson, there exists a significant disparity between the two. Except for the fact that the TFFs $f_+^{B\to\eta}(q^2)$ and $f_+^{B\to\eta^{\prime}}(q^2)$ are closer in shape than the case of $f_+^{D\to\eta^{(\prime)}}(q^2)$, there are other two reasons for this discrepancy, e.g.
\begin{itemize}
  \item[1)] The whole physical region $q^2$ corresponding to them is quite different. The larger interval can be fitted with two similar parameters, whereas the smaller interval cannot be fitted by two similar parameters.
  \item[2)] According to the SSE method employed, if two similar parameters, namely $b_1\approx b_2$, are utilized in the decay process of $D$-meson, the curve fitted exhibits significant disparities compared to the outcomes derived from our LCSR calculation. And the larger the disparity between $b_1$ and $b_2$ within the same $q^2$ region, the greater the value of $\Delta$, resulting in a more pronounced steepness of the fitted curve.
\end{itemize}
In addition to extrapolating with the SSE, there are other fitting methods, which have also been suggested in the literature, e.g.
\begin{itemize}
  \item[1)] In early studies of beauty and charm semileptonic decays, the form of TFFs were commonly assumed to adhere to the simple pole model (also referred to as nearest pole dominance)~\cite{Aleksan:1994bh}.
  \begin{align}
  f_+^{H\to\eta^{(')}}(q^2)=\frac{f_+^{H\to\eta^{(')}}(0)}{1-q^2/m^2_{H^*}}.
  \end{align}
  Here $m_H$ and $m_{H^*}$ represent the $B(D)$-meson mass and the corresponding vector meson resonance, respectively. This model oversimplifies the actual dynamics, and the two free parameters ($f_+^{H\to\eta^{(')}}(0)$ and $m_{H^*}$) do not fit the experimental data well when $q^2$ range is large.

  \item[2)] Becirevic-Kaidalov (BK) parameterization~\cite{Becirevic:1999kt}:
  \begin{align}
  f_+^{H\to\eta^{(')}}(q^2)=\frac{f_+^{H\to\eta^{(')}}(0)}{(1-q^2/m^2_{H^*})(1-\alpha_{\rm BK}q^2/m^2_{H})}.
  \end{align}
  where $\alpha_{\rm BK}$ is a free parameter. This parameterization of TFFs for heavy-to-light decay conforms to the heavy quark scaling law and avoids the introduction of explicit ``dipole" forms, which has been used in the analyses of systematically lattice data and experimental studies of the   semileptonic TFFs.

  \begin{figure*}[htb!]
\begin{center}
\includegraphics[width=0.24\textwidth]{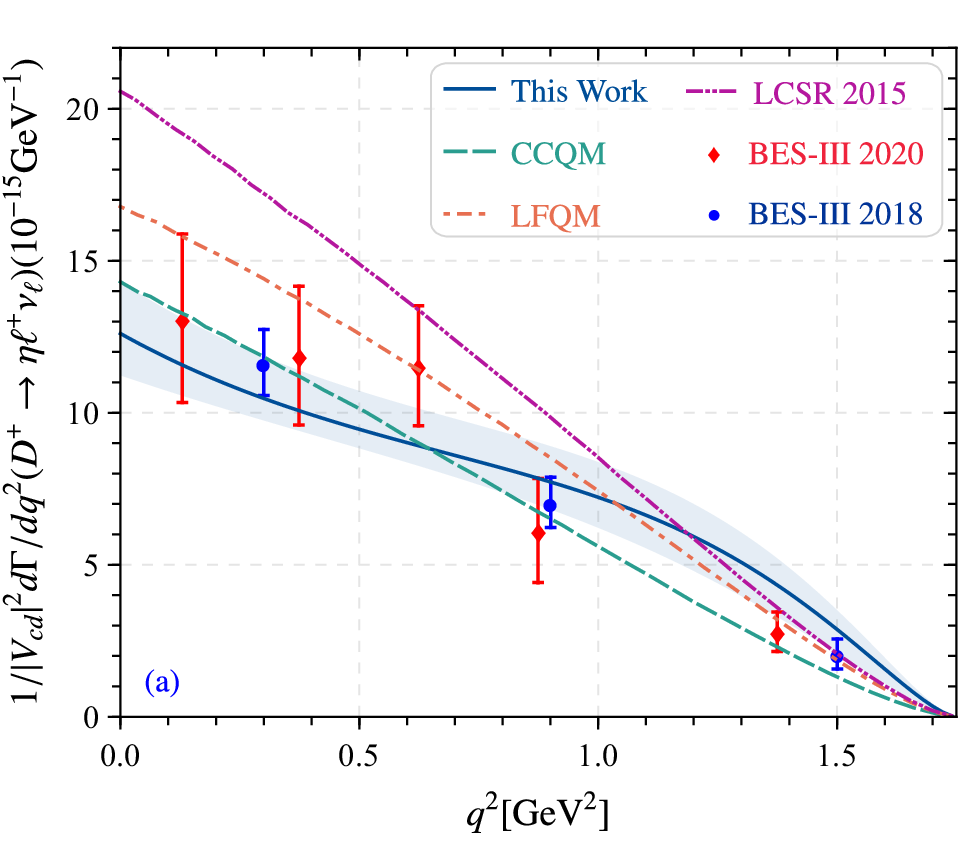}
\includegraphics[width=0.235\textwidth]{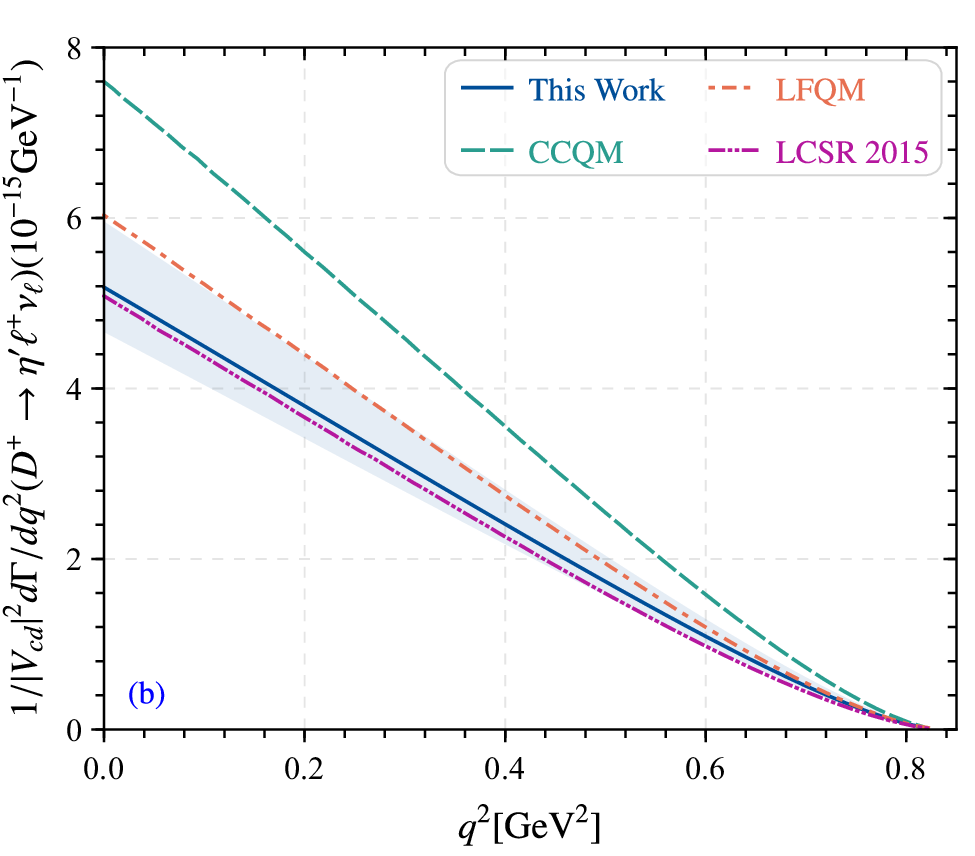}
\includegraphics[width=0.24\textwidth]{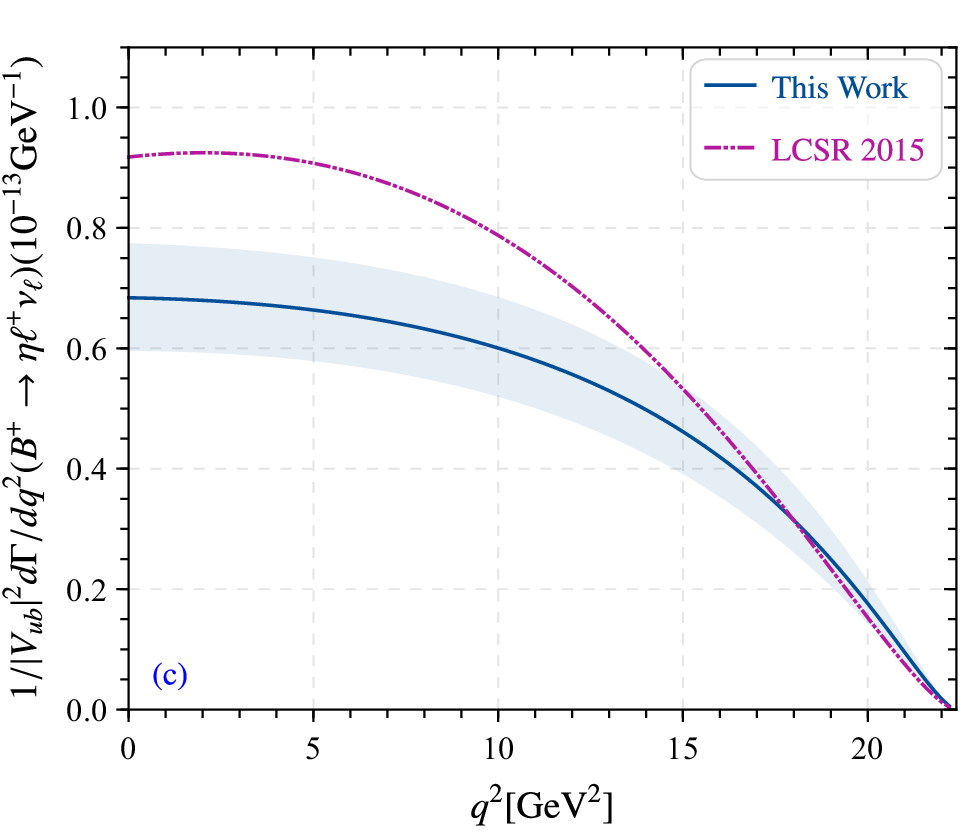}
\includegraphics[width=0.24\textwidth]{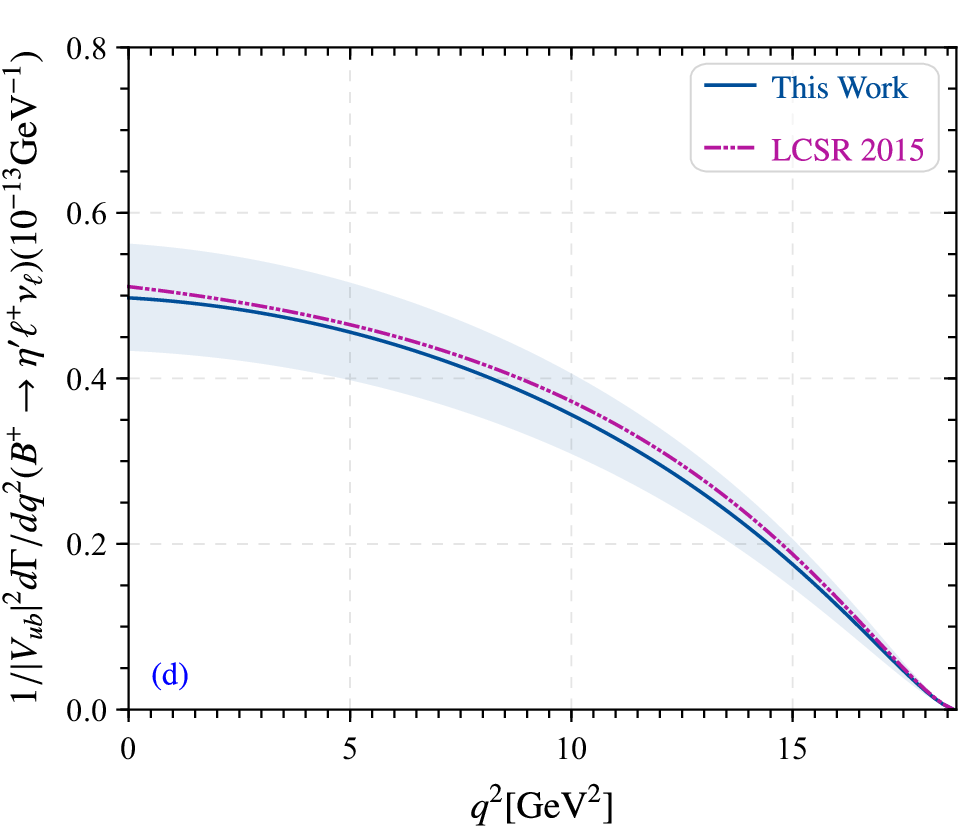}
\end{center}
\caption{(Color online) Differential decay widthes for $B(D)^+\to{\eta^{(\prime)}}{{\ell}^{+}}\nu_\ell$ in whole $q^2$-region, where the solid line is the central value and the shaded band shows its uncertainty. As a comparison, the predictions using different theoretical approaches and the experimental data, such as CCQM~\cite{Ivanov:2019nqd}, LFQM~\cite{Verma:2011yw}, LCSR~\cite{Duplancic:2015zna} and BESIII collaboration~\cite{Ablikim:2020hsc, BESIII:2018eom}, are also presented. }
\label{fig:width}
\end{figure*}

  \item[3)] Ball-Zwicky (BZ) parameterization, commonly referred to as the double-pole parameterization~\cite{Ball:2001fp}.
  \begin{align}
  f_+^{H\to\eta^{(')}}(q^2)=\frac{f_+^{H\to\eta^{(')}}(0)}{1-\alpha(q^2/m^2_{H})+\beta(q^2/m^2_{H})^2}.
  \end{align}
  This parameterization is valid in the entire kinematical range of semileptonic decays and is consistent with vector-meson dominance at large momentum transfer. Besides, the parameters $\alpha$ and $\beta$ are rather sensitive to the chosen range for $q^2$ in the actual calculation.

  \item[4)] The Boyd-Grinstein-Lebed (BGL) parametrization is based on the dispersion relation to describe the heavy mesons semileptonic decay form factors, independent of heavy quark symmetry~\cite{Boyd:1995sq}.
  \begin{align}
  f_+^{H\to\eta^{(')}}(q^2)=\frac{1}{P(t)\phi(t)}\sum_{k=0}^\infty a_k[z(t,t_0)]^k,
  \end{align}
  where $P(t)$ is Blaschke factor that depends on the masses of the sub-threshold resonances. The coefficients $a_k$ are unknown constants constrained to obey $\sum_{k=0}^\infty |a_k|^2\leq 1$. The BGL parameterization is often referred to as the series expansion (SE), whose starting point is to extend the TFFs defined in the physical range (from $q^2=0$ to $(m_H-m_{\eta^{(')}})^2$) to analytic functions throughout the complex $t=q^2$ plane. By selecting an appropriate normalization function $\phi(t)$, a simple dispersion bounds for SE coefficients can be obtained~\cite{Bharucha:2010im}.
\end{itemize}

All the aforementioned parameterization methods can be utilized for the extrapolation of the TFFs, each with their own advantages and disadvantages~\cite{Ball:2006jz, Dalgic:2006dt}. The central value of the TFFs $f_+^{H\to\eta^{(\prime)}}(q^2)$ are obtained through various parameterization methods is illustrated in Fig.~\ref{fig:fit}. The TFFs $f_+^{H\to\eta'}(q^2)$ obtained by different parameterization methods exhibit consistency in across the entire $q^2$ region, but show discrepancy in $f_+^{H\to\eta}(q^2)$ at the larger $q^2$ region. Particularly for $f_+^{D\to\eta}(q^2)$, the behaviors derived from the four fitting methods differ significantly, while the trends of SSE and SE are relatively similar. The simple pole model is deemed overly simplistic, thus this portion of the graph is omitted.

Opting for SSE parameterization have a significant advantage as it effectively translates the near-threshold behavior of the TFFs into a constraining condition on the expansion coefficient. The extrapolated TFFs in whole $q^2$-region are given in Fig.~\ref{fig:fq}, where some typical theoretical and experimental results are presented as a comparison, such as CCQM~\cite{Ivanov:2019nqd}, LFQM~\cite{Verma:2011yw}, LCSR 2015~\cite{Duplancic:2015zna}, pQCD~\cite{Charng:2006zj} and BESIII 2020~\cite{Ablikim:2020hsc}. The solid lines in Fig.~\ref{fig:fq} denote the center values of the LCSR predictions, where the shaded areas are theoretical uncertainties from all the mentioned error sources. The thicker shaded bands represent the LCSR predictions, which have been extrapolated to physically allowable $q^2$-region. Fig.~\ref{fig:fq} indicates that: 1) Our present LCSR prediction of $f_+^{D\to\eta}(q^2)$ is in good agreement with BESIII data~\cite{Ablikim:2020hsc}; 2) Our present LCSR prediction of $f_+^{D\to\eta'}(q^2)$ is consistent with the LFQM prediction~\cite{Verma:2011yw} and the LCSR 2015~\cite{Duplancic:2015zna} predictions within errors; 3) Our present LCSR predictions of $f_+^{B\to\eta^{(\prime)}}(q^2)$ are close to the LCSR 2015 prediction~\cite{Duplancic:2015zna}, and their values at $q^2= 0$ are consistent with the pQCD prediction~\cite{Charng:2006zj} within errors. In Fig.~\ref{fig:fq}, the upward trend for the $f_+^{D\to\eta}(q^2)$ TFF exhibits a significant difference in the large $q^2$ region with other theoretical results and experimental data, which there are two reasons for this difference, the main reason is because we get a smaller $f^{D\to\eta}_{+}(q^2)$ TFF at $q^2=0$. In order to obtain a smooth fitting curve, there is an overall upward trend in the large $q^2$ region. And the secondary reason is the utilization of distinct methods for extrapolation. The other theoretical groups in Fig~\ref{fig:fq} employed the double-pole fitting method to obtain the form factor across the entire physical region, whereas we opted for the simpler SSE approach.

Fig.\ref{fig:width} presents the differential decay widthes for $B(D)^+\to{\eta^{(\prime)}}{{\ell}^{+}}\nu_\ell$ without CKM matrix elements. As a comparison, the predictions using different theoretical approaches and the experimental data, such as CCQM~\cite{Ivanov:2019nqd}, LFQM~\cite{Verma:2011yw}, LCSR~\cite{Duplancic:2015zna} and BESIII collaboration~\cite{Ablikim:2020hsc, BESIII:2018eom}, are also presented. The differential decay width $d\Gamma/|V_{\rm cd}|dq^2 (D^+\to\eta {\ell^ + }{\nu _\ell})$ agrees with the BESIII 2018~\cite{BESIII:2018eom} and BESIII 2020~\cite{Ablikim:2020hsc} within errors.

\begin{table}[htb]
\renewcommand\arraystretch{1.3}
\center
\small
\caption{CKM matrix elements $|V_{cd}|$ and $|V_{ub}|$ are obtained by different experimental groups according to different decay processes.}\label{Tab:CKM}
\begin{tabular}{l l l l}
\hline
Mode ~~~~~~~~~~~~& Channels~~~~~~~~~~~~~~ & $|V_{cd}|$         \\       \hline
This work        & $D^{+}\to\eta e^{+}\nu_e$         & $0.236_{-0.017}^{+0.017}$ \\
This work        & $D^{+}\to\eta \mu^{+}\nu_{\mu}$    & $0.228_{-0.017}^{+0.017}$  \\
This work        & $D^{+}\to\eta' e^{+}\nu_e$         & $0.253_{-0.032}^{+0.028}$  \\
BESIII 2020~\cite{Ablikim:2020hsc}    & $D^{+}\to\eta \mu^{+}\nu_{\mu}$      & $0.242\pm0.028\pm0.033$   \\
BESIII 2013~\cite{BESIII:2013iro}     & $D^+\to \mu^+ \nu_{\mu}$       & $0.221\pm0.006\pm0.005$   \\
BaBar 2014~\cite{BaBar:2014xzf}       & $D^0\to\pi^- e^+ \nu_e$      & $0.206\pm0.007\pm0.009$   \\
CLEO 2009~\cite{CLEO:2009svp}            & $D\to \pi e^+\nu_{e}$          & $0.234\pm0.009\pm0.025$  \\
HFLAV~\cite{HFLAV:2019otj}          & $D\to \pi\ell\nu_{\ell} $      & $0.225\pm0.003\pm0.006$  \\
LQCD 2019~\cite{Lubicz:2017syv}       & $D\to \pi\ell\nu$              & $0.233\pm0.137$  \\
PDG~\cite{ParticleDataGroup:2022pth}             & $D\to \pi \ell \nu$            & $0.233\pm0.003\pm0.013$      \\ \hline
     ~~~~~~~~~~~~& Channels~~~~~~~~~~~~~~& $|V_{ub}|\times10^{-3}$     \\  \hline
This work        & $B^{+}\to\eta \ell^{+}\nu_{\ell}$   & $3.752_{-0.351}^{+0.373}$    \\
This work        & $B^{+}\to\eta' \ell^{+}\nu_{\ell}$  & $3.888_{-0.787}^{+0.688}$    \\
CLEO 2007~\cite{CLEO:2007vpk}    & $B^0\to\pi^-\ell^+\nu$  & $3.60\pm0.4\pm0.2$       \\
HFLAV~\cite{HFLAV:2019otj}       & $B\to \pi\ell\nu_{\ell}$              & $3.70\pm0.10\pm0.12$     \\
BaBar 2011~\cite{BaBar:2011xxm}    & $\bar{B}\to X_u \ell \bar{\nu}$       & $4.33\pm0.24\pm0.15$     \\
Belle~\cite{Belle:2005uxj}       & $B\to X_u \ell \nu $         & $4.09\pm0.39\pm0.18$      \\
PDG~\cite{ParticleDataGroup:2022pth}          & $B\to\pi\ell{\bar \nu}$               & $3.82\pm0.20$               \\
PDG~\cite{ParticleDataGroup:2022pth}          & $B\to X_u \ell{\bar \nu}$             & $4.13\pm0.12\pm0.18$            \\
LQCD 2015~\cite{FermilabLattice:2015mwy}         & $B\to \pi\ell\nu$           & $3.72\pm0.16$        \\ \hline
\end{tabular}
\end{table}

By matching the branching fractions and the decay lifetimes given by the PDG with the decay widthes predicted by Eq.(\ref{eq:CKM}), one may derive the CKM matrix elements $|V_{ub}|$ and $|V_{cd}|$. We put our results in Table~\ref{Tab:CKM}, where the errors are caused by all the mentioned error sources and the PDG errors for the branching fractions and the decay lifetimes. Some typical measured values of $|V_{ub}|$ and $|V_{cd}|$ are also given in Table~\ref{Tab:CKM}. The predicted $|V_{cd}|$ is within the error range of experimental result BESIII 2020. Using the fixed CKM matrix elements, our final predictions of the branching function are: ${\rm{\cal B}}(D\to\eta e {\nu_e }) = (1.11 \pm 0.07) \times {10^{ - 3}}$, ${\rm{\cal B}}(D\to\eta \mu {\nu_\mu }) = (1.04 \pm 0.11) \times {10^{ - 3}}$, ${\rm{\cal B}}(D\to\eta' e {\nu_e}) = (2.0 \pm 0.4) \times {10^{ - 4}}$, ${\rm{\cal B}}(B\to\eta \ell {\nu_\ell }) = (3.9 \pm 0.5) \times {10^{ - 5}}$, ${\rm{\cal B}}(B\to\eta' \ell {\nu_\ell }) = (2.3 \pm 0.8) \times {10^{ - 5}}$, respectively.

\section{Summary}\label{sec:summary}

In this paper, we have suggested the LCHO model (\ref{eq:LCDA}) for the $\eta_q$-meson leading-twist LCDA $\phi_{2;\eta_q}(x,\mu)$, whose moments have been calculated by using the QCD sum rules based on the QCD background field. To compare with the conventional Gegenbauer expansion for the LCDA, the LCHO model usually has better end-point behavior due to the BHL-prescription, which will be helpful to suppress the end-point singularity for the heavy-to-light meson decays. The QCD sum rules for the $0_{\rm th}$-order moment has been used to fix the $\eta_q$ decay constant, and we obtain $f_{\eta_q}=0.141\pm0.005$ GeV, which is slightly larger than the conventional value of the pion decay constant $f_{\pi}\simeq 0.130$ GeV~\cite{ParticleDataGroup:2022pth}. As an explicit application of $\phi_{2;\eta_q}$, we have calculated the TFFs $B(D)^+ \to\eta^{(\prime)}$ under the QF scheme for the $\eta-\eta'$ mixing and by using the QCD light-cone sum rules up to twist-4 accuracy and by including the next-to-leading order QCD corrections to the dominant leading-twist part. Our LCSR prediction of TFFs are consistent with most of theoretical predictions and the recent BESIII data within errors. By applying those TFFs, we get the decay widths of ${B(D)}^+\to\eta^{(\prime)}\ell^+\nu_\ell$. The magnitudes of the CKM matrix elements $|V_{\rm ub}|$ and $|V_{\rm cd}|$ have also been discussed by inversely using the PDG values for the branching fractions and the decay lifetimes. The future more precise data at the high luminosity Belle II experiment~\cite{Belle-II:2018jsg} and super tau-charm factory~\cite{Achasov:2023gey} shall be helpful to test all those results.

\hspace{2cm}

\noindent {\bf Acknowledgments:} This work was supported in part by the Chongqing Graduate Research and Innovation Foundation under Grant No.CYB23011 and No.ydstd1912, by the National Natural Science Foundation of China under Grant No.12175025, No.12265010, No.12265009 and No.12147102, the Project of Guizhou Provincial Department of Science and Technology under Grant No.ZK[2021]024 and No.ZK[2023]142, and the Project of Guizhou Provincial Department of Education under Grant No.KY[2021]030, and the Key Laboratory for Particle Physics of Guizhou Minzu University No.GZMUZK[2022]PT01.

\end{document}